\documentclass[12pt]{article}
\usepackage{epsf,epsfig,float,amssymb,latexsym}
\usepackage{graphics,psfrag}
\newcommand{\Op}{{\mathcal{O}}(p)}
\newcommand{\Opd}{{\mathcal{O}}(p^2)}
\newcommand{\Opt}{{\mathcal{O}}(p^3)}

\newcommand{\be}{\begin{equation}}
\newcommand{\ee}{\end{equation}}
\newcommand{\ba}{\begin{eqnarray}}
\newcommand{\ea}{\end{eqnarray}}

\newcommand{\nn}{\nonumber}

\newcommand{\vs}{\vspace{-0.0cm}}
\newcommand{\sigp}

\begin{document}

\thispagestyle{empty}

\vspace{2cm}

\begin{center}
{\Large{\bf On the Strangeness $-1$ S-wave Meson-Baryon Scattering}}
\end{center}
\vspace{.5cm}

\begin{center}
{\Large Jos\'e A. Oller}
\end{center}

\begin{center}
{\it  Departamento de F\'{\i}sica. Universidad de Murcia.\footnote{email: oller@um.es}\\ E-30071
Murcia.  Spain.}
\end{center}
\vspace{1cm}

\begin{abstract}
\noindent
We consider meson-baryon interactions in S-wave with strangeness $-1$. This is a non-perturbative sector
populated by plenty of resonances interacting in several two-body coupled channels.  We
study this sector combining a large set of experimental data.  The recent experiments are
 remarkably accurate demanding a
sound theoretical description to account for all the data. We employ unitary chiral
perturbation theory up to and including ${\Opd}$  to accomplish this aim. 
 The spectroscopy of our solutions is studied  within this approach, discussing  the rise 
 from the pole content of the two $\Lambda(1405)$ resonances and of 
 the  $\Lambda(1670)$, $\Lambda(1800)$, $\Sigma(1480)$, $\Sigma(1620)$ and
 $\Sigma(1750)$. We finally argue about our preferred solution.
\noindent
\end{abstract}

\newpage

\section{Introduction}
\label{sec:intro}
\def\theequation{\arabic{section}.\arabic{equation}}
\setcounter{equation}{0}

The study of strangeness $-1$ meson-baryon dynamics comprising the $\bar{K}N$ plus
coupled channels, has been renewed both from the theoretical and experimental sides.
Experimentally, we have new exciting data like the increasing improvement in the
precision of measurement of the $\alpha$ line of kaonic hydrogen accomplished recently
by DEAR \cite{DEAR}, and its foreseen better determination, with an expected error of a few eV,
by the DEAR/SIDDHARTA Collaboration \cite{sid}. This has established a
challenge to theory so as to match such precision. In this line, ref.\cite{akaki} provides 
an improvement over the traditional Deser formula for relating scattering at threshold
with the spectroscopy of hadronic atoms \cite{deser}. This  is achieved by including 
isospin breaking corrections to the Deser formula up to and including ${\cal
O}(\alpha^4,(m_u-m_d)\alpha^3)$, the traditional Deser formula being ${\cal O}(\alpha^3)$ in 
this counting, with $ \alpha$ the fine-structure constant and $m_u$, $m_d$ the masses of the lightest 
quarks $u$ and $d$. This is a first necessary step since the DEAR data have a precision of 
 20$\%$,  of the
same order as the corrections worked out in ref.\cite{akaki}. In addition, one needs a good scattering
amplitude to be implemented in this equation. The study of strangeness $-1$ has a long history 
\cite{dalitz,galileo,martin,juelich,hamaie,landau,cloudy,schat} within K-matrix models,
 dispersion relations, meson-exchange
models, quark models, cloudy bag-models or large $N_c$ QCD, just to quote a few.
 However, in more recent years it has
received a lot of attention from the application of SU(3) baryon Chiral Perturbation Theory (CHPT) to this
sector together with a unitarization procedure, see e.g., 
\cite{kaisersiegel,npa,oset,reportramos,ollerm,teamL,lutznieves,bura,opv}. Recently, ref.\cite{akaki} 
pointed out the possible inconsistency of the DEAR measurement on kaonic hydrogen
 and $K^-p$ scattering,
since the unitarized CHPT results, able to reproduce the scattering data, 
were not in agreement with DEAR. Later on, ref.\cite{bura} insisted on
this fact based on its own fits, although they only included partially the ${\Opd}$ CHPT amplitudes
\cite{reply}.
 However, the situation
changed from ref.\cite{opv} where it was shown that one can obtain fits in unitary CHPT (UCHPT),  
including full ${\Opd}$ CHPT amplitudes, which are compatible both with DEAR and with $K^-p$ scattering
data. We extend in this work the analysis of ref.\cite{opv} by including additional experimental data, recently
measured with remarkable precision by the Crystal-Ball Collaboration, for the reactions 
$K^-p\rightarrow \eta \Lambda$ \cite{nefkens} and  $\pi^0\pi^0\Sigma^0$ \cite{prakhov}. The importance of
including the latter data in any analysis of $K^-p$ interactions has been singled out in ref.\cite{magas}.
The study of $K^-p$ plus coupled channel interactions offers, from the theoretical point of view, a very
challenging test ground for chiral effective field theories of QCD since one has there plenty of
experimental data, Goldstone bosons dynamics and large and explicit SU(3) breaking. In addition, this 
sector shows a very rich spectroscopy with many I=0, 1 S-wave resonances that will be object as 
well of our study. Apart from that,
these interactions are interrelated with many other interesting areas, as listed in ref.\cite{opv}, 
e.g., possible kaon condensation in neutron-proton stars \cite{kaplan,brown,pons,starts},
 large yields of $K^-$ in heavy ions collisions 
\cite{kaos,fuchs}, kaonic atoms \cite{ramoset} or non-zero strangeness content of the proton
\cite{gzero,pavan}.

In section \ref{sec:form} we outline the theoretical formalism employed to calculate the strong
 S-wave amplitudes in
coupled channels. In section \ref{sec:a4pfits} we review the data and fits delivered 
in ref.\cite{opv} and
present an ${\Op}$ fit to the same data. In the next section  we include further data
and give new fits for  the prior and new data. These fits are classified in two families, 
particularly based on the agreement or disagreement  with respect to the DEAR measurement of kaonic hydrogen.
In section \ref{sec:spec} we discuss the pole content and its relation with
observed resonances for the most representative fits. We end with some conclusions giving reasons
 to  fix our preferred fit. 

\section{Formalism}
\label{sec:form}
\def\theequation{\arabic{section}.\arabic{equation}}
\setcounter{equation}{0}

CHPT is the effective field theory of strong interactions at low energies 
\cite{fisica,gl,ureport,u2,kaiserport,preport,eckerport}. In 
refs.\cite{gasserb,weinn} its extension to treat baryonic fields was pioneered. 
We concentrate here on processes including one baryon, both in the initial and final state, as
well as in the intermediate ones. CHPT applied to this situation is usually called  
 baryon CHPT. In the SU(2) sector it has proved very successful, see e.g. 
 \cite{ureport,u2,kaiserport}, 
 and references therein. 
 However,  due to the relatively large mass of the strange quark, pure perturbative 
 applications of SU(3) baryon CHPT suffer from converging problems. Notice that 
 while in SU(2) one has 
$m_\pi$ as an expansion parameter, for SU(3) one also has $m_K$, with $m_\pi$, $m_K$ the masses
 of pions and kaons, respectively, being the latter much larger than the former. These facts make that cancellations of large contributions at second and
third chiral order often happen with still sizable ${\cal O}(p^4)$ contributions, see, e.g.,
  refs.\cite{kaiserkn,boram,steinm}. Even more, for the case of S-wave I=0 $\bar{K}N$ 
 scattering lengths, the CHPT prediction is a disaster \cite{kaiserkn}. This is due to 
the presence of the $\Lambda(1405)$ resonance below and close to the $ \bar{K}N$ threshold.
 The situation  changes
 once the chiral expansion is implemented with a resumation of unitarity bubbles
 \cite{kaisersiegel}, showing  that chiral Lagrangians can be used in strangeness $-1$
 meson-baryon interactions reproducing this resonance. In ref.\cite{ollerm} the
  resummation of the right hand cut or unitarity cut (taking into account unitarity and analyticity)
  in the CHPT expansion was systematized to any two body process without spoiling 
 the chiral counting and the CHPT series up to the considered order. This gives rise
  to the known Unitary CHPT or UCHPT. This work originated in turn 
from a series of previous works \cite{npa,iam,nd,oset,reportramos,pin},
 where similar techniques were already
employed in meson-meson and meson-baryon  production and interactions.  
 
Meson-baryon interactions are described 
to lowest order in the CHPT expansion, i.e. at $\Op$, by
the chiral  Lagrangian
\ba
{\cal L}_1&=&\langle i\bar{B}\gamma^\mu [D_\mu,B]\rangle-m_0 
\langle\bar{B}B \rangle \nn\\
&+&\frac{D}{2}\langle \bar{B}\gamma^\mu
\gamma_5\{u_\mu,B\}\rangle +\frac{F}{2} \langle \bar{B}\gamma^\mu 
\gamma_5 [u_\mu,B]\rangle~,
\label{lag1}
\ea
where $m_0$ stands for the octet baryon mass in the SU(3) chiral limit. 
The trace $\langle \cdots\rangle$ runs over flavor indices and the 
axial-vector couplings are constrained by $F+D=g_A=1.26$. We use
 $D=0.80$ and $F=0.46$ extracted from hyperon decays \cite{hyper}. 
Furthermore,  $u_\mu=iu^\dagger
(\partial_{\mu} U) u^\dagger$, $U(\Phi)=u(\Phi)^2=\exp(i\sqrt{2}\Phi/f)$, 
with $f$ the pion decay constant in the SU(3) chiral limit, 
and the covariant derivative $D_\mu=\partial_\mu+\Gamma_\mu$ 
with $\Gamma_\mu= [u^\dagger,\partial_\mu u]/2$. 
The $3\times 3$ flavor-matrices $\Phi$ and $B$ collect the lightest
octets of pseudo-scalar mesons $(\pi,K,\eta)$ and 
baryons $(N,\Sigma,\Lambda,\Xi)$, respectively:
\ba
\Phi&=&\left(\
\begin{array}{ccc}
\frac{\pi^0}{\sqrt{2}}+\frac{\eta}{\sqrt{6}} & \pi^+ & K^+ \\
\pi^- & -\frac{\pi^0}{\sqrt{2}}+\frac{\eta}{\sqrt{6}} & K^0 \\
K^- & \bar{K}^0 &-\frac{2 \eta}{\sqrt{6}}
\end{array}\right)~,\nn\\
B&=&\left(\begin{array}{ccc}
\frac{\Sigma^0}{\sqrt{2}}+\frac{\Lambda}{\sqrt{6}} & \Sigma^+ & p \\
\Sigma^- & -\frac{\Sigma^0}{\sqrt{2}}+\frac{\Lambda}{\sqrt{6}} & n \\
\Xi^- & \Xi^0 &-\frac{2 \Lambda}{\sqrt{6}}
\end{array}\right)~.
\label{bmatrix}
\ea
 At  
next-to-leading order (NLO) in CHPT, i.e. $\Opd$,
the meson-baryon interactions are described by the Lagrangian
\ba
{\cal L}_2&=&b_0\langle \bar{B}B\rangle \langle 
\chi_+\rangle + b_D\langle{\bar{B}\{\chi_+,B\}}\rangle + b_F \langle
\bar{B}[\chi_+,B]\rangle\nn\\
&+& b_1\langle \bar{B}[u_\mu,[u^\mu,B]]\rangle + b_2 \langle
\bar{B}\{u_\mu,\{u^\mu,B\}\} \rangle \nn\\
&+&b_3\langle \bar{B}\{u_\mu,[u^\mu,B]\}\rangle
+b_4\langle \bar{B} B\rangle \langle u_\mu u^\mu\rangle+\cdots~.
\label{lag2}
\ea
Here ellipses denote terms that do not produce
new independent contributions to S-wave meson-baryon scattering at $\Opd$.
 In addition,  $\chi_+=u^\dagger \chi u^\dagger+u\chi^\dagger u$, 
$\chi=2B_0 {\cal M}_q$, ${\cal M}_q$ 
is the diagonal quark mass matrix $(m_u,m_d,m_s)$, and 
$B_0 f^2=-\langle 0| \bar{q}q|0\rangle$ the quark
 condensate in the SU(3) chiral limit. 
The $b_i$ couplings present in eq.(\ref{lag2}) are  fitted to data, with the subscript $i$ referring 
both to $B$, $D$, $F$ as well as to 1, 2, 3 ,4. Nevertheless, in the fitting process we will impose three
relations to be satisfied between the $b_i$, hence decreasing to the same extent the number
 of free parameters. 

From the Lagrangians of eqs.(\ref{lag1}) and (\ref{lag2}) we calculate the 
${\cal O}(p)$ and ${\cal O}(p^2)$ meson-baryon amplitudes. The $\Op$ expressions 
in the canonical basis for the baryons\footnote{The canonical basis is given by the 
fields $B_a$, $a=1,\ldots,8$, such that $B=\sum_{a=1}^8 B_a \lambda_a/\sqrt{2}$, with
$\lambda_a$ the Gell-Mann matrices and $B$ the matrix given in eq.(\ref{bmatrix})} are 
given in ref.\cite{ollerm}. The $\Opd$ expressions are given in ref.\cite{opv2}. 
 The calculated chiral amplitudes are then  projected in S-wave  according to, 
\be
T_{ji}(W)=\frac{1}{4\pi}\int d\Omega\, T_{ji}(W,\Omega;\sigma,\sigma)~,
\label{proj}
\ee
 where $T_{ij}(W,\Omega;\sigma,\sigma)$ is a generic meson-baryon scattering amplitude 
 of channel $i$ into channel $j$ depending
  on $W$, the total  energy in the center 
of mass frame (CM), angles $\Omega$ and the initial and final spin of the baryons, $\sigma$, 
  with 
 $\sigma=\pm 1/2$. The result of eq.(\ref{proj}) does not depend on the particular sign for 
 $\sigma$.

\begin{figure}[H]
\psfrag{a}{\begin{tabular}{l} {\small ${\Op}$}\\ {\small Seagull} \end{tabular}}
\psfrag{b}{\begin{tabular}{l} {\small ${\Op}$}\\ {\small Direct } \end{tabular}}
\psfrag{c}{\begin{tabular}{l} {\small ${\Op}$}\\ {\small Crossed } \end{tabular}}
\psfrag{d}{\begin{tabular}{l} {\small ${\Opd}$}\\ {\small Contact terms} \end{tabular}}
\centerline{\epsfig{file=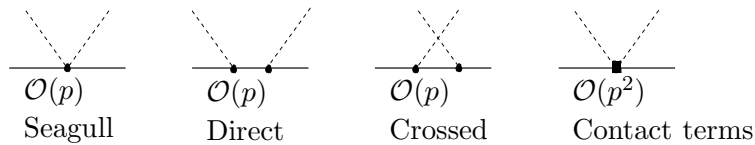,height=.6in,width=3.5in,angle=0}}
\vspace{0.2cm}
\caption[pilf]{\protect \small
 Diagrams for the calculation of the baryon CHPT scattering amplitudes up to 
 and including ${\Opd}$. The first three diagrams are ${\Op}$ while the latter is 
${\Opd}$. 
\label{fig:tree}}
\end{figure}

We have ten meson-baryon 
 coupled channels with strangeness $-1$ (or zero hypercharge): $\pi^0 \Lambda$,
$\pi^0\Sigma^0$, $\pi^-\Sigma^+$, $\pi^+ \Sigma^-$, $K^- p$, 
 $\bar{K}^0 n$, $\eta \Lambda$, $\eta
\Sigma^0$, $K^0 \Xi^0$ and $K^+ \Xi^-$, in increasing threshold 
energy order. Each channel is labelled by its position (1 to 10) 
in the previous list. 
 We denote the CHPT amplitudes at $\Op$  by 
$T_\chi^{(1)}\,\!\!_{ij}$
and  at  $\Opd$ by $T_\chi^{(2)}\,\!\!_{ij}$, with 
subscripts $ij$ indicating the scattering process $i\rightarrow j$, so that the CHPT amplitude up to
and including ${\Opd}$ is given by $T_\chi^{(1)}\,\!\!_{ij}+T_\chi^{(2)}\,\!\!_{ij}$. 
We employ these perturbative amplitudes as input for  UCHPT at NLO. The scheme 
is the following \cite{ollerm}.  Two-body partial wave amplitudes 
can be written in matrix notation as:
\be
T(W)=\left[{I}+{\cal T}(W)\cdot g(s)\right]^{-1}\cdot{\cal T}(W)~,
\label{u1}
\ee
with  $s=W^2$, the Mandelstam $s$ variable. The matrix elements
 of $T(W)$ are those of 
eq.(\ref{proj}). Eq.(\ref{u1}) was
  derived in \cite{ollerm} by employing a coupled channel dispersion relation for
  the inverse of a partial wave $T_{ij}$.
 The unitarity or right hand cut 
is taken into account easily  by the discontinuity of $T^{-1}(W)|_{ij}$ 
 for $W$ above the $i_{th}$ threshold, which is 
 given by the phase space factor $-\delta_{ij}q_i/8\pi W$, 
with $q_i$ the CM three-momentum of channel $i$. This factor is given by the imaginary part 
of the diagonal
 matrix $g(s)$, where $g(s)_i$ is the  $i_{th}$ channel unitarity bubble:
\ba
g(s)_i&=&\frac{1}{(4\pi)^2} \Bigg\{ a_i(\mu)+\log\frac{M_i^2}{\mu^2}-\frac{m_i^2-M_i^2+s}{2 s}
\log\frac{M_i^2}{m_i^2}+\frac{q_i}{W}\bigg[\log(s-\Delta+2W q_i)
   \nn \\
&+& \log(s+\Delta+2W q_i)-\log(-s+\Delta+2 W q_i)-\log(-s-\Delta+2Wq_i)\bigg] \Bigg\} ~,
\label{gs}
 \ea
 here $\Delta=m_i^2-M_i^2$ and $m_i$, $M_i$ are the 
baryon and meson masses for channel $i$, respectively. In the following, $\mu$ will be fixed to the 
value of the $\rho$ mass, $\mu=M_p\simeq 0.77$ GeV. 
 In other terms, the $g(s)_i$ satisfy a once subtracted dispersion relation,
\be
g(s)_i=g(s_0)-\frac{s-s_0}{\pi}\int_{s_{th,i}}^\infty ds'\frac{q_i}{8\pi
W'}\frac{1}{(s'-s)(s'-s_0)}~,
\ee 
whose explicit expression is given above, eq.(\ref{gs}). On the other hand, $s_{th,i}$ is 
the value of $s$ for the threshold of channel $i$. 
 The resummation 
of the right hand cut is justified in order to resum the chain of unitarity bubbles 
that is enhanced by the large masses of kaons and baryons.
This spoils the straightforward use of the chiral series \cite{weinn,reply}. 
The dispersion relation above is once subtracted 
 because 
phase space tends to a constant for $s\rightarrow \infty$. 
This is why a  subtraction constant
$a_i(\mu)$ for each channel appears in the $g(s)_i$ function. 
 In our problem, isospin symmetry reduces the number of subtraction constants
from 10 to 6 \cite{teamL}, $ a_1$, 
$a_2= a_3=a_4$, 
$ a_5= a_6$, $ a_7$, $ a_8$ 
and $ a_9= a_{10}$. 
 On the other hand,
 we keep the physical 
masses of mesons and  baryons in the calculation of 
 $g(s)_i$, which then produces pronounced cusp effects.
 The  interaction kernel ${\cal T}(W)$ (${\cal T}={\cal T}_1+{\cal T}_2+\cdots$,
 where  subscripts indicate the chiral order), 
is fixed by matching (\ref{u1}) with
 the baryon CHPT amplitudes $T_\chi$ 
order by order, as clearly explained in \cite{ollerm}. At leading order, $\Op$, 
 ${\cal T}_1=T_\chi^{(1)}$ 
while at NLO, $\Opd$, ${\cal T}_2= T_\chi^{(2)}$. The matching can be done
 to any arbitrary order and for $\Opt$ or  higher ${\cal T}_n\neq T_{\chi}^{(n)}$. 
 Explicit expression for $T_\chi^{(1)}+T_\chi^{(2)}$, and hence for 
 ${\cal T}_1+{\cal T}_2$, are given in ref.\cite{opv2}. The ${\cal T}$ matrix,
  up to and including ${\Opd}$, incorporates 
local and pole terms as well as crossed channel dynamics contributions
in the dispersion relation for $T^{-1}$, see fig.\ref{fig:tree}.

\section{Data and fits of ref.\cite{opv}}
\label{sec:a4pfits}
\def\theequation{\arabic{section}.\arabic{equation}}
\setcounter{equation}{0}

We now discuss the data employed in ref.\cite{opv} to obtain its fits $A_4^+$  and $B_4^+$,
 since we are going to use these data also in our own fits. The latter
  include the $\sigma(K^- p\rightarrow K^- p)$ elastic cross section \cite{26plb,27plb,28plb,31plb}, 
the $\sigma(K^- p\rightarrow \bar{K}^0n)$ charge exchange one
\cite{26plb,27plb,31plb,29plb,30plb}, 
 and several hyperon production reactions, 
$\sigma(K^- p\rightarrow \pi^+\Sigma^-)$ \cite{26plb,27plb,28plb}, 
$\sigma(K^- p\rightarrow \pi^-\Sigma^+)$ \cite{27plb,28plb,31plb},  
$\sigma(K^- p\rightarrow \pi^0\Sigma^0)$ \cite{27plb} 
and  $\sigma(K^- p\rightarrow \pi^0\Lambda)$ \cite{27plb}.  In our normalization the corresponding 
cross section, keeping only the S-wave, is given by
\be
\sigma(K^-p \to M B)=\frac{1}{16 \pi s}\,\frac{p'}{p}\,|T_{K^- p\to MB}|^2~,
\label{KPMB}
\ee
where $MB$ denotes the final meson-baryon system, $p'$ the final CM three-momentum and 
$p$ the initial one. 
 
In addition, we also fit the precisely measured ratios 
at the $K^- p$ threshold \cite{nowak,tovee}:
\ba
\gamma&=&\frac{\sigma(K^-p\rightarrow \pi^+\Sigma^-)}
{\sigma(K^-p\rightarrow \pi^-\Sigma^+)}=2.36\pm 0.04~,\\
R_c&=&\frac{\sigma(K^-p\rightarrow \hbox{charged particles})}
{\sigma(K^-p\rightarrow \hbox{all})}=0.664\pm0.011~,\nn\\
R_n&=&\frac{\sigma(K^-p\rightarrow \pi^0\Lambda)}
{\sigma(K^-p\rightarrow \hbox{all neutral states})}=0.189\pm 0.015.
\nn
\label{ratios}
\ea
  The first two ratios,
which are Coulomb corrected, are measured with 1.7$\%$ precision, 
 which is of the same order as  the expected isospin violations. 
Indeed, all the other observables  we fit have
uncertainties larger than $5\%$. 

 Since we are just considering the S-wave partial waves,  we  only include in the
fits those data points for the several $K^-p$ cross  sections 
with  laboratory frame   $K^-$ three-momentum  $p_K\leq 0.2$ GeV.
 This also enhances 
the sensitivity  to the lowest energy region in which  we are particularly 
interested.  We also 
 include in the fits the $\pi^{\pm}\Sigma^{\mp}$ 
event distributions from the chain of reactions $K^-p \to \Sigma^+(1660) \pi^-$, 
$\Sigma^+(1660)\to \pi^+ \Sigma \pi$  \cite{hemingway}. The $\Sigma^+ \pi^-$ and 
$\Sigma^- \pi^+$ have I=1 Clebsch-Gordan coefficients opposed in sign while both 
have the same I=0 Clebsch-Gordan coefficient. Since this process is dominated by 
the $\Lambda(1405)$ resonance, which afterwards decays into $\Sigma \pi$, we want to 
remove as much as possible the I=1 contamination. Indeed, one can observe small differences 
in the data \cite{hemingway} between the event distributions for $\Sigma^\pm \pi^\mp$ 
 due to this I=1 effect, that indeed 
is enhanced by the presence of I=1 resonances close to the $\Lambda(1405)$ energy region, 
as reported in \cite{ollerm,teamL} or within the entry $\Sigma(1480)$ 
of the PDG \cite{pdg}, qualified there as bumps.  See also ref.\cite{cosy} for a possible recent 
observation of this resonance. No I=1 resonance around the $\bar{K}N$ 
threshold is reported in refs.\cite{lutznieves,ben,lutzkolo}. We shall present our own 
results on that in the
section \ref{sec:spec}, dedicated to spectroscopy. In order to remove the interference
 with the I=1 contribution 
we take the average between the $\Sigma^\pm \pi^\mp$ event distributions.
 For the calculation  we follow \cite{ollerm}, where a generic $L=$I$=0$ source is taken for the
 generation of the final $\Sigma^\pm \pi^\mp$ particles, $L$ is the angular momentum.
 Final state interactions are taken into account in the very same way as 
for the strong meson-baryon scattering amplitudes. In this case, the ``production" vertices 
 for the $i_{th}$ channel are
the ${\cal T}_{\alpha i}$ matrix elements,  from the 
``source"$\equiv \alpha_{th}$ channel . Afterwards, final state interactions give rise 
to the factor
 $[I+K\cdot g]^{-1}$. Hence, the elementary production vertices, $R_i$, 
 because of final state interactions, change to  $R \rightarrow [I+{\cal T}]\cdot R=F$, with $F_i$
 the transition amplitude to the $i_{th}$ channel. In order to simplify matters, as
done as well in ref.\cite{ollerm}, we only consider $R_i\neq 0$ for the $\bar{K}N$ and $\pi
\Sigma$ channels, as they are the only channels with I=0 component that open in the considered energy region around the
$\Lambda(1405)$. Any other channels with I=0 component
 are much higher in energy. Hence, $R_i=(0,r,r,r,r',r',0,0,0,0)$. The final expression considered 
is then:
\be
\frac{dN_{\pi\Sigma}}{dW}=\bigg|r(D_{32}+D_{33}+D_{34})+r'(D_{35}+D_{36})\bigg|^2 p_{\pi^-\Sigma^+}~,
\label{ompis}
\ee 
with $D=\left[I+{\cal T}\cdot g\right]^{-1}$. In this way, the I=0 component 
is the only one contributing. We have taken in eq.(\ref{ompis}) the $\pi^-\Sigma^+$ channel 
 to evaluate the 
three-momentum $ p_{\pi^-\Sigma^+}$. We could have also taken the $\pi^+ \Sigma^-$, 
being the numerical effects negligible. The parameters $r$ and $r'$ are fitted to the 
average of the $\pi^\pm \Sigma^\mp$ event distribution data.

 The number of data points included in each fit,  without 
the data for the energy shift and width of kaonic hydrogen, is 97. 
 Unless the opposite is stated, we also include  in the fits 
the DEAR measurement of the shift and width of the 
$1s$ kaonic hydrogen level \cite{DEAR},
\ba
 \Delta E&=&193\pm 37 (stat) \pm 6 (syst.) \hbox{ eV},\nn\\
\Gamma&=&249 \pm 111 (stat.)\pm 39 (syst.) \hbox{ eV}\, , 
\label{deardata}
\ea 
which is around a factor two more precise than  the KEK \cite{ito}
measurement,
$\Delta E=323 \pm 63\pm 11$ eV and $\Gamma=407\pm 208 \pm 100$ eV.
To calculate the shift and width of the $1s$  
kaonic hydrogen state we use the results of \cite{akaki}
incorporating  isospin  breaking corrections up to and including 
${\cal O}(\alpha^4,(m_d-m_u)\alpha^3)$. 
The final expression taken from ref.\cite{akaki} is,
\ba
T_{KN}^{(0)}&=&4\pi\left(1+\frac{M_{K^+}}{m_p}\right)\frac{(a_0+a_1)/2+q_0 a_0
a_1}{1+q_0(a_0+a_1)/2}~,\nn\\
\Delta E-\frac{i}{2}\Gamma&=&-\frac{\alpha^3 \mu_c^3}{2\pi
M_{K^+}}\,T_{KN}^{(0)}\left\{1-\frac{\alpha \mu_c s_1(\alpha)}{4\pi
M_{K^+}}T_{KN}^{(0)}\right\}~,
\label{eqakaki}
\ea 
where, as suggested in that reference, we have taken for practical purposes $\delta T_{KN},\,
\delta_1^{vac}= 0$. The notation followed is that of ref.\cite{akaki}. We have displayed
these formulas in order to show how the strong $\bar{K}N$ scattering lengths in the isospin 
limit, $a_0$ and $a_1$
for I=0, 1,  respectively, enter in eqs.(\ref{eqakaki}). The
definition of the isospin limit is the same as in ref.\cite{akaki}, taking for the mass of the
  $K$, $\pi$ and nucleon multiplets that of the positively charged particle.  
 We compare the results obtained from eq.(\ref{eqakaki}) with those from the Deser formula 
\cite{deser}, directly given in terms of the $K^- p$ scattering length, $a_{K^-p}$, 
$\Delta E-i \,\Gamma/2=-2 \alpha^3 \mu_c^2 a_{K^- p}$, without considering the isospin limit. 
Within the uncertainties given in ref.\cite{akaki}, one can use $4\pi (1+M_{K^+}/m_p)a_{K^-p}$
 instead of $T_{KN}^{(0)}$ 
in  eq.(\ref{eqakaki}). We have checked that for all our fits 
the resulting  $\Delta E$ and $\Gamma$
are very close to those obtained directly employing eq.(\ref{eqakaki}). Hence we do not elaborate more
on this point.

  We further constrain our fits by computing   
several $\pi N$  observables calculated in baryon SU(3) CHPT 
at $\Opd$  with the values of the low energy
 constants  involved in the fit. 
Unitarity corrections in the $\pi N$ sector
are  not as large as in the 
 $S=-1$ sector, e.g., there is no something like a 
$\Lambda(1405)$ resonance close to threshold,  and hence a calculation within 
pure SU(3) baryon CHPT  is more reliable for this sector. Thus,  we calculate  at $\Opd$,    
  $a_{0+}^+$, the isospin-even pion-nucleon S-wave scattering length,
  $\sigma_{\pi N}$, the pion-nucleon $\sigma$ term, 
and $m_0$ from the value of the proton mass $m_p$,
\ba
\sigma_{\pi N}&=&-2 m_\pi^2(2b_0+b_D+b_F)~,\nn\\
a_{0+}^+&=&-\frac{m_\pi^4}{2\pi f^2}\left[(2b_0+b_D+b_F)-(b_1+b_2+b_3+2 b_4)+\frac{g_A^2}{8 m_p}\right]~,\nn\\
m_p&=&m_0-4m_K^2( b_0+ b_D-b_F)-2 m_\pi^2 (b_0+2b_F)~.
\label{opd}
\ea

\begin{figure}[H]
\psfrag{mb}{$mb$}
\psfrag{elas}{$\sigma(K^-p\to K^-p)$}
\psfrag{cexch}{$\sigma(K^-p\to \bar{K}^0 n)$}
\psfrag{pip}{$\sigma(K^-p\to \pi^+ \Sigma^-)$}
\psfrag{pim}{$\sigma(K^-p\to \pi^- \Sigma^+)$}
\psfrag{pi0}{$\sigma(K^-p\to \pi^0 \Sigma^0)$}
\psfrag{pil}{$\sigma(K^-p\to \pi^0 \Lambda)$}
\psfrag{sigpra}{\hspace{-0.7cm} \vspace{-1.5cm}$\sigma(K^-p\to \pi^0 \pi^0\Sigma^0)$}
\psfrag{nefkens}{$\sigma(K^-p\to \eta \Lambda)$}
\psfrag{hem}{$K^-p \to \Sigma^+(1660) \pi^-$}
\psfrag{pra}{\hspace{-0.1cm}$K^-p \to \pi^0\pi^0\Sigma^0$}
\psfrag{pk (MeV)}{$p_K$(MeV)}
\psfrag{pk (GeV)}{$p_K$(GeV)}
\psfrag{pis (GeV)}{$\pi \Sigma$(GeV)}
\psfrag{piss (GeV)}{$\pi \Sigma$(GeV$^2$)}
\psfrag{Event}{$\pi\Sigma$ Events}
\psfrag{Events}{$\pi^0\Sigma^0$ Events}
\centerline{\epsfig{file=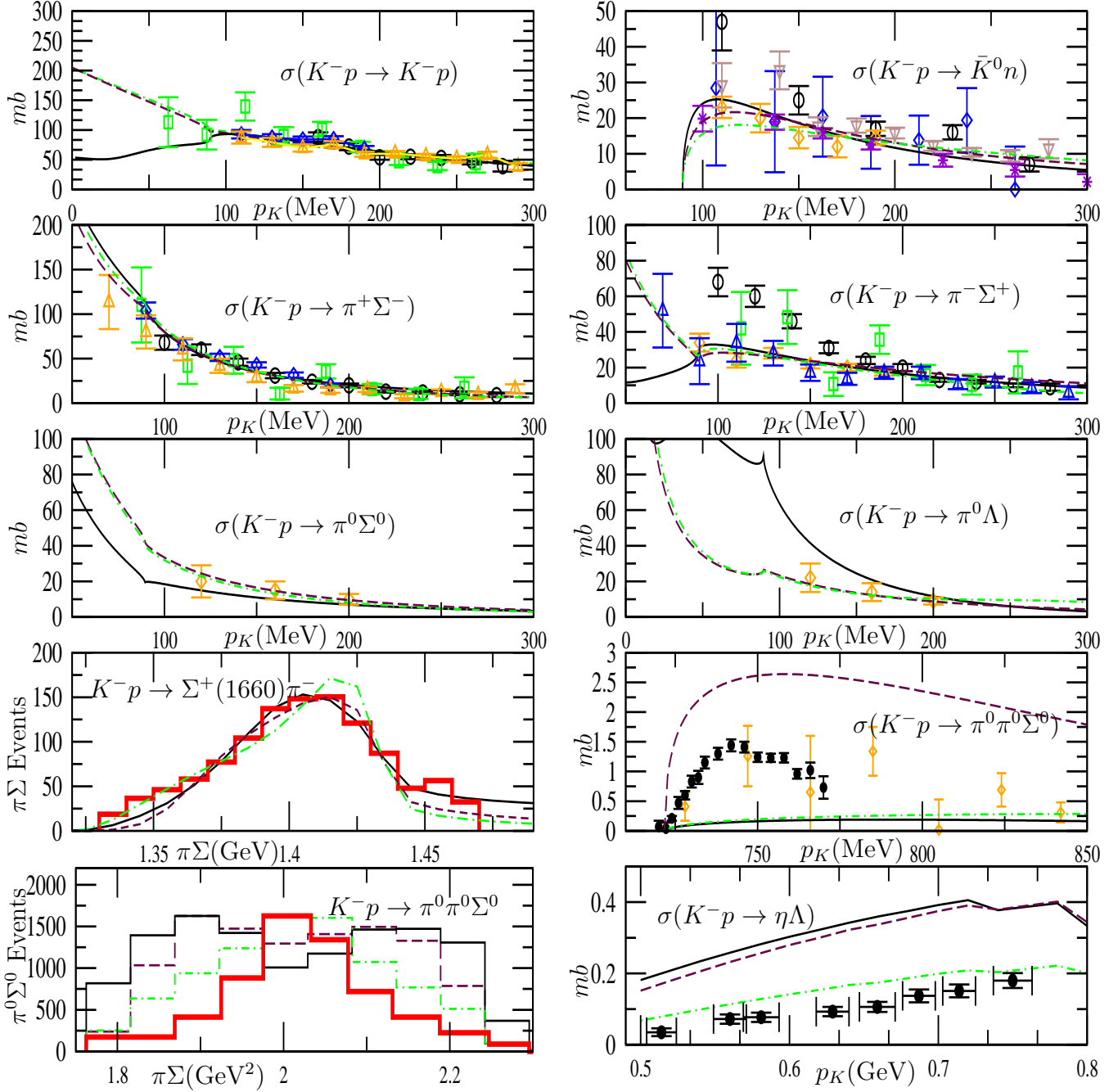,height=7.in,width=7.0in,angle=0}}
\vspace{0.2cm}
\caption[pilf]{\protect \small
 The solid lines correspond to the fit $A_4^+$, the dashed ones to  $B_4^+$ and 
 the dash-dotted lines to the ${\cal O}(p)$ fit given in table \ref{table:a4pvalues}.
  The data employed in the fit is that 
 of section \ref{sec:a4pfits}. The experimental references for the 
 first six panels are: squares \cite{26plb},
diamonds \cite{27plb}, upwards triangles \cite{28plb}, circles \cite{31plb}, stars \cite{29plb}
and downwards triangles \cite{30plb}. The data in the seventh panel, thick solid line,
 are from ref.\cite{hemingway}. The 
 circles and diamonds in the eighth panel are from refs.\cite{nefkens,baxter}, in order.
The thick solid line in the ninth panel is the experimental data \cite{prakhov}. While 
 the data of the tenth panel correspond to ref.\cite{prakhov}. 
\label{fig:a4p}}
\end{figure} 

  We do not   consider the isospin-odd $\pi N$ 
scattering length $a_{0+}^-$  since at this order 
is just given by $g_A$, in good agreement with experiment \cite{ulf}.
The $\sigma_{\pi N}$ term receives sizable higher order corrections 
from the  mesonic cloud  which are expected to be positive 
 and around $10$ MeV \cite{gasser2}.
 Since we evaluate  it just at $\Opd$, we enforce 
$\sigma_{\pi N}=$ 20, 30 or 40 MeV in the fits ($\sigma_{\pi N}=45\pm 8 $ MeV \cite{glsigma}). 
For the same reason,  $m_0=$  0.7 or 0.8 GeV was enforced in ref.\cite{opv} 
($m_0=0.77\pm 0.11$ GeV from ref.\cite{boram} or $0.71  \lesssim  m_0 \lesssim 1.07$ GeV
 \cite{frink}). In the new fits to be discussed in the next section,  we 
 use  $m_0=0.9\pm 0.2$, as suggested by ref.\cite{frink}. 
We also include the value 
$a_{0+}^+=-(1\pm 1) \cdot 10^{-2}\,m_\pi^{-1}$ in the fit procedure. 
This value  results after considering its experimental measurement \cite{schroder}, 
 $a_{0+}^+=-(0.25\pm 0.49) \cdot 10^{-2}\,m_\pi^{-1}$, 
and the theoretical expectation of
  positive $\Opt$ corrections around $+1\cdot 10^{-2}\,m_\pi^{-1}$ 
  from unitarity \cite{ulf}.  Thus, the inclusion of eq.(\ref{opd}) implies 
three relations between the $b_i$ that basically reduce by three\footnote{Here ``basically" 
is because $a_{0^+}^+$ and $m_0$ are given with some error, while $\sigma_{\pi N}$ is 
fixed.} the number of fitted parameters 
 shown in tables \ref{table:a4pvalues}, \ref{table:a4pnewvalues} and \ref{table:b4newvalues}. 
  It is worth stressing that for all the fits we minimize strictly the $\chi^2$, 
that is, each data point is   weighted according to 
its experimental error. We do not include the data
 from ref.\cite{31plb} in the $\sigma(K^- p\rightarrow \pi^-\Sigma^+)$ 
cross section since they are incompatible with all the  other  data.

\begin{table}[ht]
\begin{center}
\begin{tabular}{|l|l|r|r|r|}
\hline
Units&    &  $A_4^+$   &  $B_4^+$  &   ${\Op}$  \\
  \hline
MeV &         $f$  & $79.8$   & $89.2$     & 88.0  \\
GeV$^{-1}$ & $b_0$ & $-0.855$ & $-0.318$  & $0^*$    \\
GeV$^{-1}$ & $b_D$ & $+0.715$  &$-0.101$    & $0^*$  \\
GeV$^{-1}$ & $b_F$ & $-0.036$  &$-0.314$   & $0^*$        \\
GeV$^{-1}$ & $b_1$ & $+0.605$  &$-0.193$   & $0^*$ \\
GeV$^{-1}$ & $b_2$ & $+1.075$  & $-0.275$ & $0^*$   \\
GeV$^{-1}$ & $b_3$ & $-0.189$  & $-0.153$  & $0^*$    \\
GeV$^{-1}$ & $b_4$ & $-1.249$ & $-0.277$   & $0^*$ \\
	  & $a_1$  & $-1.155$ & $-1.570$  & $-0.472$\\
	  & $a_2$  & $-0.383$ & $-2.062$  & $-1.572$\\
	  & $a_5$  & $-1.304$ & $-2.605$  & $-1.266$\\
	  & $a_7$  & $-1.519$ & $-1.568$  & $-1.853$\\
	  & $a_8$  & $-1.212$ & $-2.064$  & $-1.210$\\
	  & $a_9$  & $-0.145$ & $-0.886$  & $+3.337$\\
 \hline
\end{tabular}
\caption{Resulting values for the parameters of the fits $A_4^+$, third column,
 and $B_4^+$, fourth column. The $\Op$ fit is given in the fifth column. The fits $A_4^+$ 
and $B_4^+$ are from ref.\cite{opv}.
In the last column the asterisks mean that the corresponding parameters are fixed to $0$ since 
is an $\Op$ calculation.
 \label{table:a4pvalues}}
\end{center}
\end{table}

In fig.\ref{fig:a4p} we show the scattering data and the $\pi \Sigma$ event distributions 
in the first seven panels, from left to right and top to bottom. The solid 
and dashed curves, correspond to the $A_4^+$ and $B_4^+$ fits, respectively. They  
reproduce well the data included in 
the fits of ref.\cite{opv} and discussed in this section. The last three
 panels in the same figure correspond to  
other data not considered in the present fits nor in ref.\cite{opv}. 
In the eighth panel we show the
total cross section $\sigma(K^- p\rightarrow \eta \Lambda)$. The solid points come from 
ref.\cite{nefkens} while the diamonds are much older \cite{baxter}. The
  $\pi^0 \Sigma^0$ event distribution and the 
total cross section for the reaction $K^- p\rightarrow \pi^0\pi^0\Sigma^0$, measured in 
\cite{prakhov},  are displayed, respectively, in the ninth
 and tenth panels. The data in the last three panels will be presented and 
discussed further in the next section.  It is clear from the figure that the $A_4^+$
 and $B_4^+$ fits
 do not reproduce adequately the data in the last three panels. 
  In fig.\ref{fig:a4p}, we also show by the dash-dotted lines 
the $\Op$ fit to the same data. 
 As we see, this fit, with 4 free parameters less than the others,\footnote{We recall that in the
  ${\Opd}$ fits we also consider 
$m_0$, $\sigma_{\pi N}$ and $a_{0^+}^+$, directly given by eqs.(\ref{opd}) in terms of the 
low energy constants $b_i$. In this way, although there 
are 7 low energy constant only four are really kept
 as free parameters in the ${\Opd}$ fits. This statement, however, is only approximate because 
 neither $m_0$ nor $a_{0^+}^+$ are exactly sticked to a value, 
 $m_0=0.9\pm 0.2$ GeV and $a_{0^+}^+=-(1\pm 1) 10^{-2}$.} 
is able to reproduce the scattering data  but fails as well in the reproduction 
of $\sigma(K^-p\to \eta \Lambda)$, although its disagreement  with the data from 
the reaction $K^-p \to \pi^0\pi^0\Sigma^0$ \cite{prakhov} is neatly smaller than for the fits 
$A_4^+$ and $B_4^+$. In table \ref{table:a4pvalues} we give the values of the fitted parameters. 
We  show
 in  table \ref{table:a4pratios} the resulting values 
 for the ratios of eq.(\ref{ratios}) and observe that for all the fits there is agreement 
 with  experiment for $\gamma$ and 
 $R_n$ within the small errors given. For $R_c$, the fits  $B_4^+$ and ${\Op}$ agree within the experimental error, while 
  $A_4^+$ agrees with the experimental value  at the 
 level of $5\%$, which is equally satisfactory since we do not intend at this stage 
 to arrive at such 
 precision in the description of strong interactions in this sector, where even isospin breaking
 corrections should be systematically included.
  We also show in the same table the kaonic hydrogen data included
in the fit, as well as  other magnitudes as explained in the table caption. 
 Only the $A_4^+$ fit is in agreement with the shift and width of kaonic
hydrogen from DEAR \cite{DEAR}. The fits $B_4^+$ and $\Op$ are in
agreement with KEK \cite{ito} but disagrees with DEAR \cite{DEAR}. We also show the calculated 
energy shift and width of kaonic hydrogen from the Deser formula. The differences 
with respect to the results from the more elaborated eq.(\ref{eqakaki}) are huge for $\Gamma$ in 
the fits $B_4^+$ and ${\Op}$ and by far much  smaller, a correction of just a few percent, 
 in the fit $A_4^+$.  For the $\Op$ fit the value for $a_9$ is quite large, although one has to keep in mind
 that this parameter has very large errors as given by the minimizing subroutine \cite{minuit}. 
 Indeed,  its upper error is much larger than the value of the parameter itself. Hence one 
 concludes that this parameter is left undetermined by the $\Op$ fit. We should also remark 
 that all the parameters in table \ref{table:a4pvalues} are of natural size. The $b_i$ are of 
order GeV$^{-1}$ and the $a_i$ of around 1. This is the natural size for the $a_i$ since 
from the value of the imaginary part of $g(s)_i$ above threshold, $-q_i/8\pi W$, multiplying 
it by $16 \pi^2$, the prefactor in eq.(\ref{gs}), one has $-q_i/ 2 \pi W$. Taking for $W$ the mass of a 
nucleon, $m_N$, the ratio is then the quotient 
of $q_i$ over $\simeq 150$ MeV, which is typically a quantity of order 
1. Furthermore, from the unitarity corrections to the chiral series induced by $g(s)$, which 
start at $\Opt$, one can derive the scale $\Lambda_U\simeq 16 \pi^2 f^2/2 m_N |a_i|$. Again 
this scale is of natural size, around the mass of the $\rho$, for $|a_i|=1$. However, 
for larger values of $|a_i|$ it can be quite small, e.g., of the order of the difference 
between the masses of the nucleon and  $\Delta(1232)$. Regarding the precise values for 
$b_0$, $b_D$ and $b_F$ we can compare our values in table \ref{table:a4pvalues}
 with the determination based on resonance saturation 
and reproduction of the masses of the lightest baryon octet
 and $\sigma_{\pi N}$ from ref.\cite{boram}. The authors of ref.\cite{boram} conclude that 
 $b_0\simeq -0.61$, $b_D \simeq 0.08$ and $b_F\simeq -0.32$ in units of GeV$^{-1}$. From a pure 
 $\Opd$ analysis of baryon masses and $\sigma_{\pi N}$ one has, in the same units,
  $b_D=0.064$, $b_F=-0.209$ and  $b_0=-0.518$ or $-0.807$ depending on whether the value for 
  $\sigma_{\pi N}$ is taken from ref.\cite{glsigma} or from ref.\cite{pavan}, respectively. 
These values look 
 somewhat closer to those of the fit $B_4^+$ than to the values of the fit $A_4^+$. 
 However, the comparison is not straightforward since we  employ the $\Opd$ couplings
  in UCHPT, which  resums large contributions in this sector, 
 so that  there is no reason why the values should be the same as in CHPT. It is  
 remarkable that the value for $b_3$ is very similar both in the fits $A_4^+$ and $B_4^+$. 
 Indeed, from ref.\cite{boram} one also has $b_1=-0.004$, $b_2=+0.018$, $b_3=-0.187$ and 
 $b_4=-0.109$, hence the value for $b_3$ is quite similar also to those in the table. 
 Finally, in ref.\cite{kaisersiegel} a value of around $-0.15$ GeV$^{-1}$ was given as well  
 and in many of the fits of ref.\cite{bura} values around $-0.2$ GeV$^{-1}$ are reported. 
 In the fits that we present later, $b_3$ mostly appear between $-0.2$ and $-0.3$ GeV$^{-1}$.

\begin{table}[ht]
\begin{center}
\begin{tabular}{|c|r|r|r|r|}
\hline
 & $A_4^+$    & $B_4^+$    &  $\Op$ \\
\hline
$\gamma$  & 2.36 & 2.36 & 2.35 \\
$R_c$   & 0.628 & 0.655&  0.667 \\
$R_n$   & 0.172 & 0.195&  0.205 \\
$\Delta E$~(eV) & 201 & 403 & 390 \\   
$\Gamma$~(eV) &  338 & 477& 525   \\
$\Delta E_D$~(eV) & 209 & 416 & 394 \\   
$\Gamma_D$~(eV)  &  346 & 662& 716   \\
$a_{K^-p}$ (fm)& $-0.51+i\,0.42$   & $-1.01+i\,0.80$  & $-0.96+i\,0.87$  \\
$a_0$      (fm)& $-1.23+i\,0.45$   & $-1.63+i\,0.81$  & $-1.55+i\,0.87$ \\
$a_1$      (fm)& $0.98+i\,0.35$   & $-0.01+i\,0.54$  & $-0.03+i\,0.65$    \\
$\delta_{\pi \Lambda}(\Xi)$  ($^\circ$) & $2.5$ & $0.2$ & $-1.9$ \\
$m_0$  (GeV) & $0.8^*$ & $0.8^*$& $\ldots$\\
$a_{0+}^+$  ($10^{-2}\cdot M_\pi^{-1}$)& $-1.2$  & $-1.7$ & $\ldots$\\
$\sigma_{\pi N}$ (MeV) & $40^*$ & $40^*$ & $\ldots$\\
\hline
\hline
\end{tabular}
\caption{Values for the ratios at threshold of  eq.(\ref{ratios}), energy shift ($\Delta E$) 
and width ($\Gamma$) of the ground state of the kaonic hydrogen, both using eq.(\ref{eqakaki}) and 
the Deser formula, the latter indicated by the subscript $D$. We also give the  $K^-p$ S-wave scattering length, $a_{K^-p}$,  
I=0 and $1$ $\bar{K}N$ scattering lengths in the isospin limit, $a_0$ and $a_1$, respectively, 
the difference between the P- and S-wave $\pi \Lambda$ phase shifts at the $\Xi^-$ mass, 
$\delta_{\pi \Lambda}(\Xi)$, the lightest baryon mass in the
chiral limit, $m_0$, the isospin even $\pi N$ scattering length, $a_{0+}^+$, and the 
enforced $\sigma_{\pi N}=40^*$ MeV  for the ${\Opd}$ fits. 
\label{table:a4pratios}}
\end{center}
\end{table}

   The commented discrepancy of
the $A_4^+$ and $B_4^+$ fits with the data not included in ref.\cite{opv}, corresponding to the last three
panels of fig.\ref{fig:a4p}, leads us to consider new
fits that include these new data from the beginning.

\section{New fits with additional recent data }
\label{sec:newfits}
\def\theequation{\arabic{section}.\arabic{equation}}
\setcounter{equation}{0}
In addition to the data set described in section \ref{sec:a4pfits}, we now include in the fits
 the following data, already shown in the last three panels of fig.\ref{fig:a4p}:

\begin{itemize}
\item[i)] The $\sigma(K^-p \to \eta \Lambda)$ cross section  was measured accurately in 
ref.\cite{nefkens} from threshold
 up to around $p_K=770$ MeV ($\sqrt{s}=1.69$ GeV), with $p_K$ the
 kaon three-momentum in the laboratory frame. These are 17 data points with 
 small  error bars as shown by the circles in the eighth panel of the result figures, namely, 
 figs.\ref{fig:a4p}, \ref{fig:a4pnew}, \ref{fig:b4new}. We also consider older data
  on this reaction \cite{baxter} from $p_K=728$ up to $934$ MeV ($\sqrt{s}=1.76$ GeV). They are 
much less precise than the previous data and up to $p_K=0.85$ GeV are shown in the 
 eighth panel of the same figures. Both data sets include a total of 29 new points.

\item[ii)] Data from the reaction $K^- p\rightarrow \Sigma^0\pi^0\pi^0$ recently measured in 
ref.\cite{prakhov}. These data comprise 
the $\pi^0 \Sigma^0$ event distribution, shown in the ninth panel of the result figures 
 by the thick solid line, and measurements of the associated total cross section, given
by the circles in the tenth panel of the same figures. The measurement of the cross
 section is quite accurate, as one can see from 
the figure, with $p_K$ from 0.5 GeV up to 0.75 GeV. The 
error bars given are calculated from ref.\cite{prakhov} by adding in quadrature the statistical
 errors (explicitly
given in the paper) and a systematic error of $10\%$ (the upper bound estimated in this reference for this
source of error). These data constitute 18 demanding  new points.\footnote{I
 warmly acknowledge E. Oset for having stressed to me the importance of these new data.}  

\item[iii)] Finally,  we also include the recently measured difference between the P- 
and S-wave $\pi \Lambda$ phase shifts at the $\Xi^-$ mass, from  the 
determination of the  $\Xi^-\rightarrow \Lambda \pi^-$   
decay parameters. The results are 
$\delta_P-\delta_S=(4.6\pm 1.4\pm 1.2)^{\hbox{\small o}}$ 
 \cite{hyperCP} and  
$(3.2\pm 5.3\pm 0.7 )^{\hbox{\small o}}$ \cite{e756}.
 Neglecting  the tiny P-wave phase shift \cite{valencia}, this quantity just corresponds to minus 
 $\delta_S$. As already given in ref.\cite{opv}, we obtain 
 for this quantity  $2.5^\circ$ 
for the fit $A_4^+$ and $ 0.2^\circ$ for the fit $B_4^+$. For the $\Op$ fit one has 
$-1.9^\circ$, see table \ref{table:a4pratios}. 
Hence, the fit $A_4^+$ is the only one in agreement with the measurement 
at the level of one $\sigma$. In the following we denote by $ \delta_{\pi \Lambda}(\Xi)$ this 
phase shift difference.
\end{itemize}

Thus, in total we have 153 ``scattering" data points while in ref.\cite{opv} the number of 
 ``scattering" data points, 97, was significantly smaller.

We follow a similar strategy as in the fits of \cite{opv} and then
 consider fits constrained to give 
 $\sigma_{\pi N}=20$, 30 and 40 MeV. On the other hand, $m_0=0.9\pm0.2$ is included in the fits, 
where the range of values  is taken from ref.\cite{frink}, and 
is compatible as well with that of ref.\cite{boram}. Other works on baryon masses from baryon CHPT,
 in some  or other variant, are \cite{lebed,holsteinm0,sherer,lutzm0}.

The reaction $K^-p \rightarrow \eta \Lambda$, accurately measured 
by the Crystal Ball Collaboration 
\cite{nefkens}, was also considered in refs.\cite{manley,ben,nieves},
 where it was assumed to proceed in S-wave. This assumption is well suited
  since the data from ref.\cite{nefkens} is  close
 to threshold and hence S-wave should dominate, this is also indicated 
  by the angular distributions \cite{nefkens}.  
We follow here this assumption as well and thus the strong $K^-p\rightarrow \eta\Lambda$ amplitude 
will be taken in S-wave.  According to our
normalization we have,
\be
\sigma(K^-p\rightarrow \eta \Lambda)=\frac{1}{16\pi \, s}\,\frac{p'}{p}\,
\left|T_{K^-p\rightarrow \eta\Lambda}\right|^2~,
\ee
as in eq.(\ref{KPMB}), 
with $p'$ the CM three momentum of the  $\eta \Lambda$ system and $p$ that of the initial $K^-p $ state.

\begin{figure}[H]
\psfrag{p}{$p$}
\psfrag{si}{$\Sigma^0$}
\psfrag{pi}{$\pi^0$}
\psfrag{K}{$K^-$}
\centerline{\epsfig{file=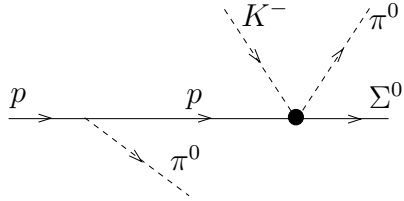,height=1.0in,width=2.0in,angle=0}}
\vspace{0.2cm}
\caption[pilf]{\protect \small
Production process for the $K^- p\rightarrow \pi^0 \pi^0 \Sigma^0$ reaction.
\label{fig:magas}}
\end{figure}

For the calculation of the $\Sigma^0\pi^0$ event distribution and the total cross section of 
the reaction 
$K^- p\rightarrow \pi^0 \pi^0 \Sigma^0 $, we follow the scheme of
ref.\cite{magas}, although we use fully relativistic amplitudes. In ref.\cite{magas} several
 production
mechanisms for the final $\pi^0\pi^0\Sigma^0$ state are included and discussed in connection 
to the 
related process $\pi^- p \rightarrow K^0 \pi \Sigma$, studied in ref.\cite{vicente}. Interestingly, all of
them are negligible compared with the diagram shown in fig.\ref{fig:magas}. The thick dot at the right 
of the figure means that the $full$ $K^- p\rightarrow \pi^0 \Sigma^0$ S-wave 
is used. Here we are assuming
that the process is dominated by the S-wave meson-baryon amplitude, which is justified since we are close to
the threshold of the reaction, see the last panel of figs.\ref{fig:a4p}, \ref{fig:a4pnew} or \ref{fig:b4new}.
 This diagram is so much enhanced
compared with other possible ones \cite{magas} 
due to the almost on-shell character of the intermediate proton. As said above, we have recalculated this diagram
in a fully Lorentz covariant way, as also done with the interaction kernel for our S-wave
 amplitudes. 
Numerically these relativistic corrections do not affect appreciably the results as compared with 
the non-relativistic limit taken in ref.\cite{magas}. Had the emitted meson  been a kaon, 
things would have been different, since then large factors of $M_K/m_p$ would have appeared.
The finding of ref.\cite{magas}, concerning the dominance of  the diagram
 of fig.\ref{fig:magas}  compared with any other considered mechanism,
  makes us confident
 about the reliability of the approach  and, hence, we include this reaction in our data set. 

Our final expression for the reaction $K^-p\rightarrow \pi^0 \pi^0 \Sigma^0$ is:
\ba
t_{\beta \alpha}&=&\frac{D+F}{2f}\frac{i\,A_p A_Q}{(p-q_1)^2 - m_p^2}\chi_\beta \left\{
\vec{p}\vec{\sigma}\left[ \frac{q_1^0}{(E_p+m_p)}+\frac{q_1^0}{E_Q+m_p}+
\frac{\vec{q}_1^2-2\vec{p}\,\vec{q}_1}{(E_p+m_p)(E_Q+m_p)}\right] \right. \nn\\
 &-&\left. \vec{q}_1\vec{\sigma}\left[ 1+\frac{q_1^0}{E_Q+m_p}-\frac{\vec{p}^2}{(E_p+m_p)(E_Q+m_p)}\right] 
 T_{K^-p\to \pi^0\Sigma^0}(W_{13})
+(q_1\leftrightarrow q_2) \right\}\chi_\alpha~.
\label{eqpra}
\ea

In the previous equation, $Q=p-q_1$ and $p$, $q_1$ and $q_2$ are the four-momenta of the incoming
 proton and outgoing
pions, respectively, and $W_{13}$ is CM energy of the $\Sigma^0$ and the second pion. 
The subscripts $\alpha$ and $\beta$ refer to the spins of the proton
 and $\Sigma^0$, in order.
  The $\chi_i$ are Pauli spinors, $A_p=\sqrt{m_p+E_p}$ and $A_Q=\sqrt{m_p+E_Q}$, with $E_q$  
the proton energy for three-momentum $\vec{q}$, 
$E_q=\sqrt{m_p^2+\vec{q}^2}$. The exchange $(q_1\leftrightarrow q_2)$ in the end of
 eq.(\ref{eqpra}) guarantees the indistinguishableness of the two emitted neutral pions.
  This is the source of a major
 background for the $\Lambda(1405)$ resonance shape in the event distribution that 
 makes the $\Lambda(1405)$ resonance to appear wider. Taking into account the phase
 space for the three final particles we have the following expression for the cross section:
\ba
\sigma(K^- p\rightarrow \pi^0\pi^0 \Sigma^0)=
\frac{1/2}{1024\, s \,\pi^5}\int d\cos\theta_{p'}\, d\phi_{p'}\, d\phi_3\, dm_{23}^2 \,dm_{12}^2
\,\frac{1}{2}\!\!\sum_{\alpha,\beta=1}^2|t_{\alpha \beta}|^2~,
\label{4.3}
\ea 
where $p'$ is the four-momentum of the $\Sigma^0$, $\phi_3$ the azimuthal angle of the second pion,
$m_{12}$ the invariant mass of the $\Sigma^0$ and the first $\pi^0$, while $m_{23}$ is that of the two pions.
 The symmetry factor $1/2$ for the calculation of the total cross sections, due to the identical neutral
 pions, is explicitly shown. In the calculation of the event distribution it should be removed, but since
 the latter is not normalized we show it for both calculations. Incidentally, we also mention
that the S-wave amplitude appearing in eq.(\ref{eqpra}) is evaluated in the $K^-$-(intermediate $p$) CM frame,
which is not the CM of the whole process, in which is expressed eq.(\ref{4.3}).
 We have worked out 
the corresponding Wigner rotation matrices to calculate the scattering amplitudes in the global CM frame 
from the S-wave in the $K^-$-(intermediate $p$) CM frame. But, since their numerical effects are negligible,
 we refrain from including them and  giving further details about their calculation.

\subsection{New $A$ type fits}
\label{subsec:anew}

\begin{figure}[H]
\psfrag{mb}{$mb$}
\psfrag{elas}{$\sigma(K^-p\to K^-p)$}
\psfrag{cexch}{$\sigma(K^-p\to \bar{K}^0 n)$}
\psfrag{pip}{$\sigma(K^-p\to \pi^+ \Sigma^-)$}
\psfrag{pim}{$\sigma(K^-p\to \pi^- \Sigma^+)$}
\psfrag{pi0}{$\sigma(K^-p\to \pi^0 \Sigma^0)$}
\psfrag{pil}{$\sigma(K^-p\to \pi^0 \Lambda)$}
\psfrag{sigpra}{\hspace{-0.7cm} \vspace{-1.5cm}$\sigma(K^-p\to \pi^0 \pi^0\Sigma^0)$}
\psfrag{nefkens}{$\sigma(K^-p\to \eta \Lambda)$}
\psfrag{hem}{$K^-p \to \Sigma^+(1660) \pi^-$}
\psfrag{pra}{\hspace{-0.1cm}$K^-p \to \pi^0\pi^0\Sigma^0$}
\psfrag{pk (MeV)}{$p_K$(MeV)}
\psfrag{pk (GeV)}{$p_K$(GeV)}
\psfrag{pis (GeV)}{$\pi \Sigma$(GeV)}
\psfrag{piss (GeV)}{$\pi \Sigma$(GeV$^2$)}
\psfrag{Event}{$\pi\Sigma$ Events}
\psfrag{Events}{$\pi^0\Sigma^0$ Events}
\centerline{\epsfig{file=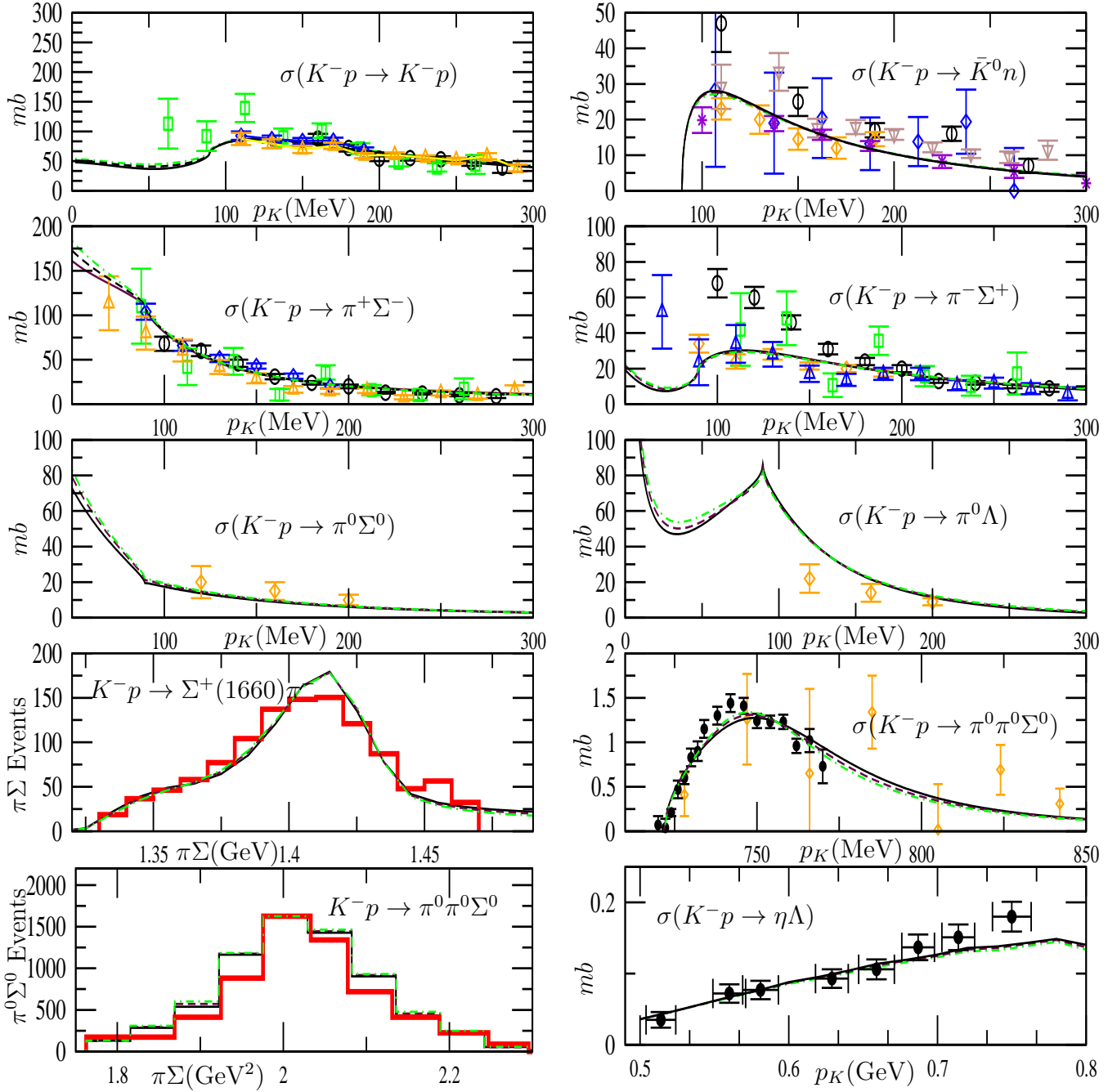,height=7.0in,width=7.0in,angle=0}}
\vspace{0.2cm}
\caption[pilf]{\protect \small
 The  solid lines correspond to the $\sigma=40^*$ MeV fit,
the dashed lines  to the $30^*$ MeV fit, and the
 dash-dotted curves to the $20^*$ MeV one of table 
 \ref{table:a4pnewvalues}. The different lines can be barely distinguished. For experimental references
 see the caption of fig.\ref{fig:a4p}. 
\label{fig:a4pnew}}
\end{figure}

We first discuss those fits that reproduce the DEAR accurate measurement, eq.(\ref{deardata}), 
of the width and shift of kaonic hydrogen, together with the rest of data. We show in 
fig.\ref{fig:a4pnew} the reproduction of the scattering data for these 
new fits, that include the additional data discussed in this section. We distinguish the
 fits according to the enforced $\sigma_{\pi N}$ value introduced in the fit and calculated from
  eq.(\ref{opd}). These values, in MeV, are  $40^*$ (solid), $30^*$ (dashed) and $20^*$ (dash-dotted lines). 
 These fits, once the new data is included, originate from the $A_4^+$ one, and many fitted
values of  the parameters, shown in table \ref{table:a4pnewvalues}, are quite similar to those of $A_4^+$ 
 given in table \ref{table:a4pvalues}. The main difference is the value for $a_7$ concerning the
 $\eta \Lambda$ channel, that
is much smaller now than it was in table \ref{table:a4pvalues}.
The value for $f$ is also a few MeV smaller now than for $A_4^+$. We have also tested
 another fit with $\sigma_{\pi N}=45\pm 18$ MeV, taking for $\sigma_{\pi N}$ 
 the central value from ref.\cite{glsigma} and adding linearly the error given in this reference and 
the expected uncertainties from higher orders \cite{gasser2}. Nonetheless, the resulting fit is somehow
intermediate between the fits with $\sigma_{\pi N}=20^*$ and $30^*$ and we do not consider it
any further. We obtain a good reproduction of the scattering data as shown in fig.\ref{fig:a4pnew}. 
Although the different lines in this figure can be barely distinguishable, we show the different fits
separately in table \ref{table:a4pnewvalues} to illustrate how different fits can  give rather similar
results. The main differences in the outputs, as shown in table \ref{table:a4newratios}, 
come from the values  of $m_0$ and, of course, of the enforced 
$\sigma_{\pi N}$. In addition, 
 we also give in this table several other observables as in table \ref{table:a4pratios}. 
 For the ratio $R_c$, the values given in table \ref{table:a4pnewvalues} agree with the experiment, 
 eq.(\ref{ratios}), within $5\%$, like in the
 case of  $A_4^+$. It is worth stressing the perfect agreement 
 with DEAR, concerning the energy shift and width for kaonic hydrogen, for all the fits of table 
 \ref{table:a4pnewvalues}, while, at the same time,  
all the scattering data shown in fig.\ref{fig:a4pnew}, plus $\delta_{\pi \Lambda}(\Xi)$, 
$m_0$, $\sigma_{\pi N}$ and $a_{0+}^+$, are reproduced too.

The $a_{K^-p}$ scattering lengths shown in the previous table are similar to those of 
the fit $A_4^+$ in table \ref{table:a4pratios}. Of course, they are much smaller in absolute value 
than those of fits $B_4^+$ and ${\Op}$. This is related to the fact that the new fits, as $A_4^+$,
 reproduce the DEAR data which, by the Deser formula, requires a much smaller 
 scattering length than those of the fits $B_4^+$ and ${\Op}$. 
Let us recall that the $A_4^+$ fit of ref.\cite{opv}  was the first chiral 
 fit to be in agreement with the 
 recent and accurate kaonic
hydrogen data from ref.\cite{DEAR} and the scattering data of section 
\ref{sec:a4pfits}. Nonetheless, since the fits in table \ref{table:a4pnewvalues}
are also able to provide a good reproduction
 of the new precise data from refs.\cite{nefkens,prakhov},
  they are preferred by us over the $A_4^+$ one.

It is remarkable as well the agreement
 with the measurement of $\delta_{\pi \Lambda}(\Xi)$ from refs.\cite{hyperCP,e756}. 
  The values in table \ref{table:a4newratios} are considerably
 larger than those obtained in ref.\cite{ollerm2} from an ${\Op}$ analysis using UCHPT, where
 the range $-1.1^\circ \lesssim \delta_{\pi\Lambda}\lesssim 0^\circ$ 
 was determined from an analysis of the scattering data of 
 section \ref{sec:a4pfits}.  Hence we see that the effect of the higher orders
in the kernel ${\cal T}$ are quite relevant for a precise determination of this quantity.

    
\begin{table}[ht]
\begin{center}
\begin{tabular}{|l|l|r|r|r|}
\hline
Units& $\sigma_{\pi N}$ & $20^*$ & $30^*$ & $40^*$  \\
     & MeV           &     &      &   \\
\hline
MeV &         $f$      & $75.2$    & $71.8$    & $67.8$   \\
GeV$^{-1}$ & $b_0$     & $-0.615$  & $-0.750$  & $-0.884$    \\
GeV$^{-1}$ & $b_D$     & $+0.818$  & $+0.848$  & $+0.873$   \\
GeV$^{-1}$ & $b_F$     & $-0.114$  & $-0.130$  & $-0.138$        \\
GeV$^{-1}$ & $b_1$     & $+0.660$  & $+0.670$   & $+0.676$   \\
GeV$^{-1}$ & $b_2$     & $+1.144$  & $+1.169$   & $+1.189$   \\
GeV$^{-1}$ & $b_3$     & $-0.297$  & $-0.316$  & $-0.315$    \\
GeV$^{-1}$ & $b_4$     & $-1.048$  & $-1.181$  & $-1.307$  \\
	  & $a_1$      & $-1.786$  & $-1.591$  & $-1.413$  \\
	  & $a_2$      & $-0.519$  & $-0.454$  & $-0.386$  \\
	  & $a_5$      & $-1.185$  & $-1.170$  & $-1.156$  \\
	  & $a_7$      & $-5.251$  & $-5.209$  & $-5.123$  \\
	  & $a_8$      & $-1.316$  & $-1.310$  & $-1.308$  \\
	  & $a_9$      & $-1.186$  & $-1.132$  & $-1.050$  \\
 \hline
\end{tabular}
\caption{Fits, presented in 
subsection \ref{subsec:anew}, that agree with the DEAR data, eq.(\ref{deardata}). The  
$\sigma_{\pi N}$ value enforced in the fits is given in the 
first row.
 \label{table:a4pnewvalues}}
\end{center}
\end{table}

\begin{table}[H]
\begin{center}
\begin{tabular}{|c|r|r|r|r|}
\hline
$\sigma_{\pi N}$ & $20^*$    & $30^*$    &  $40^*$ \\
\hline
$\gamma$  & 2.36 & 2.36 & 2.37 \\
$R_c$   & 0.629 & 0.628&  0.628 \\
$R_n$   & 0.168 & 0.171&  0.173 \\
$\Delta E$ (eV) & 194 & 192 & 192 \\   
$\Gamma$  (eV)&  324 & 302& 270   \\
$\Delta E_D$ (eV) & 204 & 204 & 207 \\   
$\Gamma_D$  (eV)&  361 & 338& 305   \\
$a_{K^-p}$ (fm)& $-0.49+i\,0.44 $   & $-0.49+i\,0.41$  & $-0.50+i\,0.37$  \\
$a_0$      (fm)& $-1.07+i\,0.53 $   & $-1.04+i\,0.50$  & $-1.02+i\,0.45$ \\
$a_1$      (fm)& $0.44+i\,0.15 $   & $0.40+i\,0.15$  & $0.33+i\,0.14$    \\
$\delta_{\pi \Lambda}(\Xi)$  ($^\circ$) & 3.4& 4.5 & 5.7\\
$m_0$  (GeV) & 1.2 & 1.1& 1.0\\
$a_{0+}^+$  ($10^{-2}\cdot M_\pi^{-1}$)& $-2.0$  & $-2.2$ & $-2.2$\\
\hline
\end{tabular}
\caption{Fits, given in table \ref{table:a4pnewvalues}, that agree with the the DEAR data, eq.(\ref{deardata}).
 The $\sigma_{\pi N}$ value enforced in the fits is given in the 
first row The notation is like in table \ref{table:a4pratios}.
\label{table:a4newratios}}
\end{center}
\end{table}


\subsection{New $B$ type fits}
\label{subsec:bnew}

\begin{table}[ht]
\begin{center}
\begin{tabular}{|l|l|r|r|r|r|}
\hline
Units& $\sigma_{\pi N}$ & $20^*$ & $30^*$ & $40^*$  & ${\Op}$ \\
     & MeV           &     &      &    &\\
\hline
MeV &         $f$      & $95.8$    & $113.2$  & $100.0$    & 93.9  \\
GeV$^{-1}$ & $b_0$     & $-0.201$  & $-0.159$  & $-0.487$  &  $0^*$  \\
GeV$^{-1}$ & $b_D$     & $-0.005$  & $-0.297$  & $0.127$   & $0^*$\\
GeV$^{-1}$ & $b_F$     & $-0.133$  & $-0.157$  & $-0.188$  & $0^*$   \\
GeV$^{-1}$ & $b_1$     & $+0.122$  & $+0.016$   & $+0.135$ & $0^*$\\
GeV$^{-1}$ & $b_2$     & $-0.080$  & $-0.151$   & $-0.037$ & $0^*$ \\
GeV$^{-1}$ & $b_3$     & $-0.533$  & $-0.281$  & $-0.494$  & $0^*$ \\
GeV$^{-1}$ & $b_4$     & $+0.028$  & $-0.291$  & $-0.173$  & $0^*$\\
	  & $a_1$      & $+4.037$  & $+4.188$  & $+2.930$  & $-2.958$\\
	  & $a_2$      & $-2.063$  & $-3.129$  & $-2.400$  & $-1.479$\\
	  & $a_5$      & $-1.131$  & $-1.214 $  & $-1.225$ & $-1.330$\\
	  & $a_7$      & $-3.488$  & $-3.000$  & $-2.795$  & $-1.805$\\
	  & $a_8$      & $-0.347$  & $+0.642 $  & $+2.906$ & $-0.655$\\
	  & $a_9$      & $-1.767$  & $-2.109$  & $-1.913$  & $-1.918$\\
 \hline
\end{tabular}
\caption{Fits, discussed in subsection \ref{subsec:bnew}, that do not agree with the DEAR data,
eq.(\ref{deardata}).
 The enforced $ \sigma_{\pi N}$ value in the fit is shown in the first line.
 \label{table:b4newvalues}}
\end{center}
\end{table}

\begin{table}[H]
\begin{center}
\begin{tabular}{|c|r|r|r|r|r|}
\hline
$\sigma_{\pi N}$ & $20^*$    & $30^*$    &  $40^*$ & ${\Op}$\\
\hline
$\gamma$  & 2.34 & 2.35 & 2.34 & 2.32 \\
$R_c$   & 0.643 & 0.643&  0.644 & 0.637 \\
$R_n$   & 0.160 & 0.163&  0.176 & 0.193\\
$\Delta E$ (eV) & 436 & 409 & 450 & 348\\   
$\Gamma$  (eV)  & 614    & 681 & 591 & 611 \\
$\Delta E_D$ (eV) & 418 & 385 & 436 & 325\\   
$\Gamma_D$  (eV)  & 848 & 880 & 844 & 775 \\
$a_{K^-p}$ (fm)& $-1.01+i\,1.03 $   & $-0.93+i\,1.07$  & $-1.06+i\,1.02$  
& $-0.79+i\,0.94$ \\
$a_0$      (fm)& $-1.75+i\,1.15 $   & $-1.65+i\,1.30$  & $-1.79+i\,1.10$ 
& $-1.50+i\,1.00$\\
$a_1$      (fm)& $-0.13+i\,0.39 $   & $-0.14+i\,0.36$  & $-0.12+i\,0.46$ 
& $0.32+i\,0.46$   \\
$\delta_{\pi \Lambda}(\Xi)$  ($^\circ$) & $-1.4$ & 1.7 & $-1.2$ & $-1.4$\\
$m_0$  (GeV) & 0.8 & 0.6 & 0.7 & \ldots \\
$a_{0+}^+$  ($10^{-2}\cdot M_\pi^{-1}$)& $-0.5$  & $-1.4$ & $+0.3$ & \ldots\\
\hline
\end{tabular}
\caption{Fits, given in table \ref{table:b4newvalues}, that do not agree with the DEAR data,
eq.(\ref{deardata}).
 The notation is like in table \ref{table:a4pratios}.
\label{table:b4newratios}}
\end{center}
\end{table}

\begin{figure}[H]
\psfrag{mb}{$mb$}
\psfrag{elas}{$\sigma(K^-p\to K^-p)$}
\psfrag{cexch}{$\sigma(K^-p\to \bar{K}^0 n)$}
\psfrag{pip}{$\sigma(K^-p\to \pi^+ \Sigma^-)$}
\psfrag{pim}{$\sigma(K^-p\to \pi^- \Sigma^+)$}
\psfrag{pi0}{$\sigma(K^-p\to \pi^0 \Sigma^0)$}
\psfrag{pil}{$\sigma(K^-p\to \pi^0 \Lambda)$}
\psfrag{sigpra}{\hspace{-0.7cm} \vspace{-1.5cm}$\sigma(K^-p\to \pi^0 \pi^0\Sigma^0)$}
\psfrag{nefkens}{$\sigma(K^-p\to \eta \Lambda)$}
\psfrag{hem}{$K^-p \to \Sigma^+(1660) \pi^-$}
\psfrag{pra}{\hspace{-0.1cm}$K^-p \to \pi^0\pi^0\Sigma^0$}
\psfrag{pk (MeV)}{$p_K$(MeV)}
\psfrag{pk (GeV)}{$p_K$(GeV)}
\psfrag{pis (GeV)}{$\pi \Sigma$(GeV)}
\psfrag{piss (GeV)}{$\pi \Sigma$(GeV$^2$)}
\psfrag{Event}{$\pi\Sigma$ Events}
\psfrag{Events}{$\pi^0\Sigma^0$ Events}
\centerline{\epsfig{file=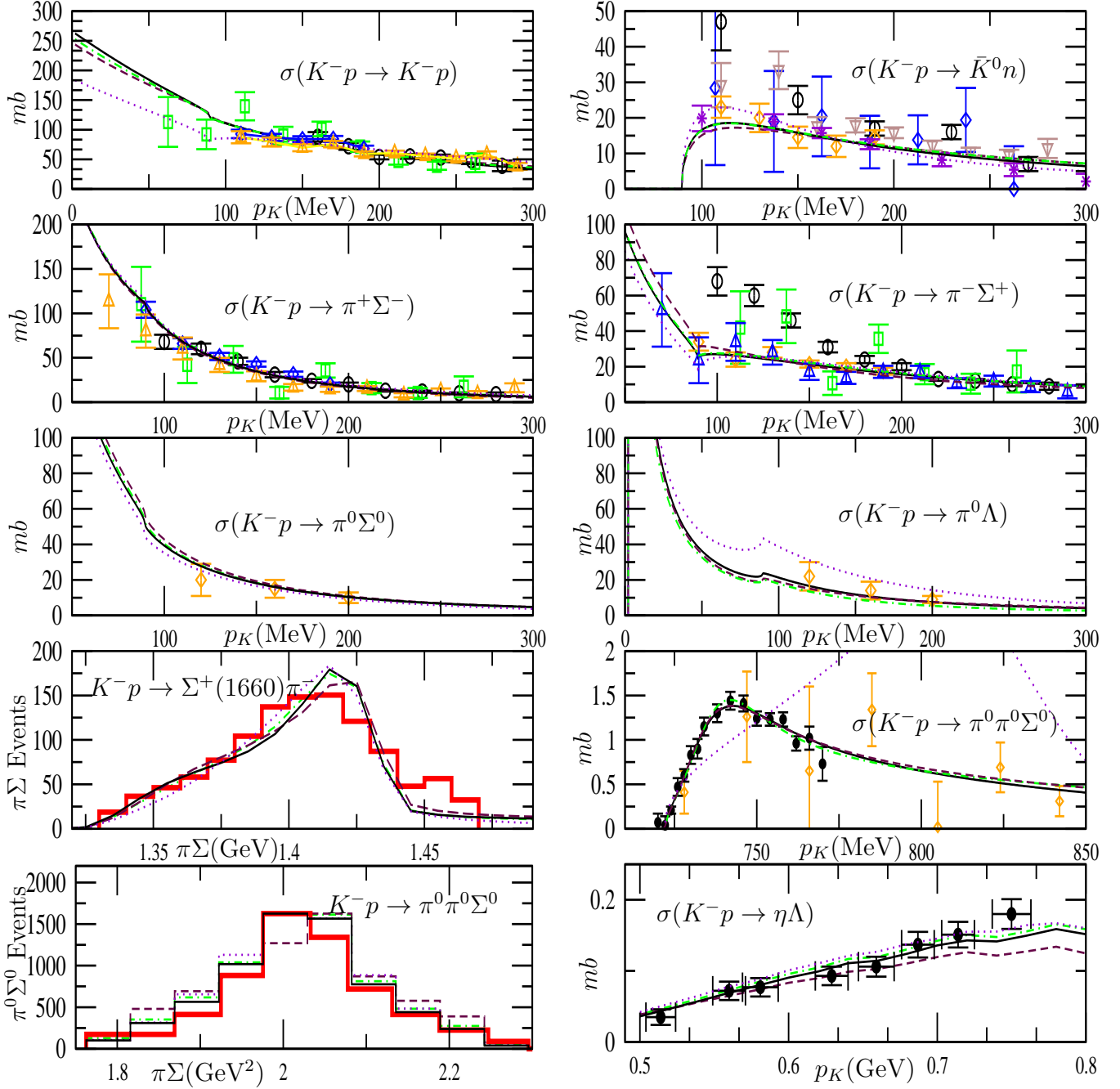,height=6.99in,width=7.0in,angle=0}}
\vspace{0.2cm}
\caption[pilf]{\protect \small
 The solid lines correspond to the $\sigma_{\pi N}=40^*$ MeV fit, the dashes lines to the
  $30^*$ MeV fit, the dash-dotted curves to the $20^*$ MeV one and the dotted lines to the ${\Op}$ 
 fit of table \ref{table:b4newvalues}.
\label{fig:b4new}}
\end{figure}

Now, we report about other fits to the whole set of data that originate from the $B_4^+$ fit of 
ref.\cite{opv}. We also include here 
an ${\Op}$ fit to all the data of this section, except for the magnitudes in eq.(\ref{opd}),
as they are defined in terms of the ${\Opd}$ couplings. Following the same scheme of presentation
 as in the prior subsection,
 we enforce in the ${\Opd}$ fits that $\sigma_{\pi N}=20^*$, $30^*$ or $40^*$ MeV. 
 The fitted parameters are given in table \ref{table:b4newvalues}, while the results
  are shown  in table \ref{table:b4newratios} and in 
  fig.\ref{fig:b4new} by the solid ($40^*$), dashed ($30^*$) and dash-dotted lines ($20^*$). 
The new ${\Op}$ fit is given in the last column of table \ref{table:b4newvalues} and its  
results are given in the last column of table \ref{table:b4newratios} and in fig.\ref{fig:b4new} 
 by the dotted lines. 


We observe that all the  fits in table \ref{table:b4newvalues} reproduce very well 
the scattering data, except for the ${\Op}$ fit which badly fails in the reproduction of 
$\sigma(K^-p\rightarrow \eta \Lambda)$, shown in the eighth panel. However, all these fits
strongly disagree with the DEAR measurement, eq.(\ref{deardata}), of the energy shift and width
of the kaonic hydrogen, particularly for the former. It is also worth noticing that for the fits 
in table \ref{table:b4newvalues} the corrections of eq.(\ref{eqakaki}) over the Deser formula for 
$\Gamma$ are large, around a $40\%$, see table \ref{table:b4newratios}, much larger than for the 
fits of table \ref{table:a4pnewvalues}. In addition, the fits $20^*$, $40^*$ and ${\Op}$ 
are also around 3 sigmas below the value of $\delta_{\pi \Lambda}(\Xi)$ measured in 
ref.\cite{hyperCP}. For the fit $40^*$ the disagreement is at the level of 2 sigmas. All this seems to
indicate that the fits of table \ref{table:a4pnewvalues} give a better overall reproduction
of the data on $\bar{K}N$ scattering than those in table \ref{table:b4newvalues}. Of course, the 
 hypothetical confirmation of the DEAR data on the energy shift and width of kaonic hydrogen  
by  the DEAR/SIDDHARTA Collaboration \cite{sid} will certainly 
 refute the fits in table \ref{table:b4newvalues}.

\section{Spectroscopy}
\label{sec:spec}
\def\theequation{\arabic{section}.\arabic{equation}}
\setcounter{equation}{0}
In this section we discuss in detail the pole content of our main fit, the third one in table
\ref{table:a4pnewvalues}. This fit will be referred in the following as I. 
We also present more briefly those poles corresponding to the $40^*$ and 
${\Op}$ fits of table \ref{table:b4newvalues}. The former will be called in the subsequent as 
II. Those other fits in  tables \ref{table:a4pnewvalues} and \ref{table:b4newvalues}  
have a pole content very similar to that of the considered ${\Opd}$ fits, I and II, respectively,
 and hence, we will not discuss them  separately for the sake of brevity. 

We only consider those Riemann sheets that are connected continuously to the physical sheet 
in some energy region of the physical axis.  The physical Riemann sheet is such that the imaginary 
part of the modulus of the
three-momentum associated with every channel is positive. The other Riemann sheets are
defined depending on which three-momenta are evaluated in the other sheet of the square root, with
an additional minus sign. 
The first non-physical Riemann sheet, 1RS, is reached when crossing the physical axis between the thresholds
 of $\pi \Lambda$ and $\pi \Sigma$, from 1.25 to 1.33 GeV, approximately.\footnote{In the definition of the sheets we just talk about 
 the $ \bar{K}N$, $\pi \Sigma$ or $K\Xi$ thresholds, although in the physical case, because of 
 isospin violation, these ``thresholds" indeed refer to a  narrow region, less than 10 MeV wide 
 at most.} 
   The so called second sheet, 2RS, is reached when crossing the physical axis between the
thresholds of $\pi \Sigma$ and $\bar{K}N$, around 1.34 and 1.43 GeV, respectively. The 
third sheet, 3RS, is connected continuously to the physical sheet between the thresholds of $\bar{K}N$ and 
$\eta \Lambda$, 1.44 and 1.66 GeV, approximately. The fourth sheet, 4RS, can be reached when crossing the 
physical axis between the $\eta \Lambda$ and $\eta \Sigma$ thresholds, from around 1.66 to 1.74 GeV. 
The fifth sheet, 5RS, is connected to the physical one between the thresholds of
$\eta \Sigma$ and $K\Xi$, around 1.74 and 1.81 GeV, respectively. And finally, the sixth sheet, 6RS, is reached by
crossing the physical axis above the $K\Xi$ threshold, approximately at 1.81 GeV. In all the 
sheets, NRS, one has $\hbox{Im} p_j\leq 0$, for $j\leq N$, and 
 $\hbox{Im} p_j\geq 0$ for $j>N$. (For $N=6$ one must understand that all the three-momenta have
 negative imaginary part.)

Once the pole position is known, one  can then calculate the couplings by
performing the limit, 
\be
T_{ij}=\lim_{s\to s_R}-\frac{g_i g_j}{s-s_R}~,
\ee
with $s_R$ the pole position  for the s Mandelstam variable. The
$g_i$ is the coupling of the pole to the channel $i_{th}$.

\begin{figure}[H]
\centerline{\epsfig{file=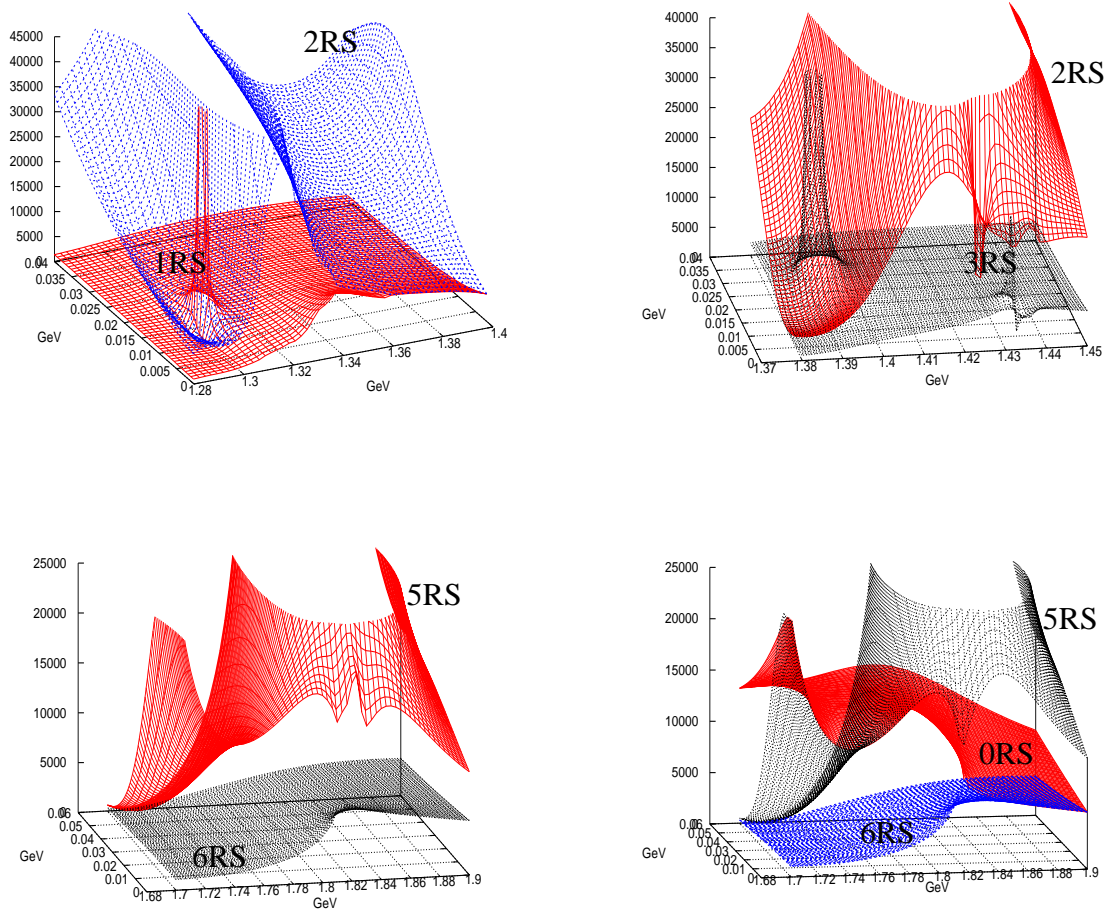,height=5.6in,width=6.5in,angle=0}}
\vspace{0.2cm}
\caption[pilf]{\protect \small
Some poles with I=0 for the fit I. From left to right and top to bottom, we have the square 
modulus of the I=0 
S-waves: $\pi\Sigma$, $\bar{K}N$, $K\Xi$ and of $K\Xi$ in the isospin limit. 
\label{fig:a4pnewpoles0}}
\end{figure} 

\begin{table}[ht]
\begin{center}
\begin{tabular}{|r|r|r|r|r|r|r|r|r|r|}
\hline
Re(Pole) & -Im(Pole) &Sheet & & & & & & & \\
 $|g_{\pi\Lambda}|$ & $|g_{\pi\Sigma}|_0$ & $|g_{\pi\Sigma}|_1$ 
 & $|g_{\pi\Sigma}|_2$ & $|g_{\bar{K} N}|_0$ & $|g_{\bar{K} N}|_1$ & $|g_{\eta\Lambda}|$ 
 & $|g_{\eta\Sigma}|$ & $|g_{K\Xi}|_0$ & $|g_{K\Xi}|_1$\\
\hline
1301 & 13 & 1RS & & & & & & &\\
0.03 & 1.12 & 0.02 & 0.01 & 5.83 & 0.05 & 0.41 & 0.04 & 2.11 & 0.03 \\
\hline
1309 & 13 & 2RS & & & & & & &\\
0.02 & 3.66 & 0.02 & 0.02 & 4.46 & 0.04 & 0.21& 0.04 & 3.05 & 0.03 \\ 
\hline
1414 & 23 & 2RS & & & & & & &\\
0.14 & 4.24 & 0.13 & 0.01 & 4.87 & 0.39 & 0.85& 0.20 & 9.35 & 0.11 \\ 
\hline
1388 & 17 & 3RS & & & & & & &\\
0.02 & 3.81 & 0.02 & 0.02 & 1.33 & 0.04 & 0.42& 0.04 & 9.55 & 0.04 \\ 
\hline
1676 & 10 & 3RS & & & & & & &\\
0.01 & 1.28 & 0.03 & 0.00 & 1.67 & 0.01 & 2.19 & 0.07 & 5.29 & 0.07 \\ 
\hline
1673 & 18 & 4RS & & & & & & &\\
0.01 & 1.26 & 0.02 & 0.00 & 1.82 & 0.01 & 2.13 & 0.06 & 5.32 & 0.06 \\ 
\hline
1825 & 49 & 5RS & & & & & & &\\
0.02 & 2.29 & 0.02 & 0.00 & 2.10 & 0.02 & 0.89 & 0.03 & 7.43 & 0.09 \\ 
\hline
\end{tabular}
\caption{Fit I, I=0 Poles. The pole positions are given in MeV and the couplings in GeV. The couplings to
the I=1, 2 channels are always close to zero.
\label{table:a4pnewpoles0}}
\end{center}
\end{table}

\subsection{Fit I}

\begin{itemize}
\item{{\bf I=0 Poles:}}
There are two I=0 poles very close to the $\pi \Sigma$ threshold, one in the 1RS and the other
in the 2RS sheet. They are located at $1301-i\,13$ and $1309-i\,13$ MeV, for the sheets 1RS and 2RS, 
respectively. The first pole has a  small coupling\footnote{All couplings will be given in 
GeV.} 1.12 to $\pi \Sigma$, while this coupling is large for the second pole, 3.68. This makes that
the bump in the square of the $ \pi\Sigma$ I=0 amplitude is very asymmetric around the $\pi \Sigma$ 
threshold. On the left of this threshold one has the behaviour corresponding to 1RS, so that
 one observes basically a cusp
effect with very little influence of the 1RS pole, while at the right of the threshold the behaviour
is dominated by the falling at the right of the 2RS pole. This is illustrated in
the first panel of fig.\ref{fig:a4pnewpoles0}, from left to right and top to bottom, where the 
square modulus of the I=0 $\pi\Sigma$ S-wave is shown. These two poles reflect the same resonance
 because they are
connected continuously when passing softly via a continuous parameter from the 1RS to the 2RS,
 as we have checked. 
 For the $\bar{K}N$ channel  one has a peak at the 1RS pole
position, although the opening of the $\pi\Sigma$ channel makes a strong cusp effect
 that distorts strongly the resonance shape giving rise to a sharp dip 
  between the $\pi \Sigma$ thresholds along the right tail of the 2RS pole. On the 2RS
we also have another pole at $1414-i\,23$ MeV, with 
 large couplings to $\pi \Sigma$ (4.24), $\bar{K}N$ (4.87) and $K\Xi$ (9.35). Note that all the ten
coupled channels are degenerate in the SU(3) limit and hence SU(3), simply because of 
the Wigner-Eckart theorem, does permit large couplings of a resonance to much heavier channels
than the resonance mass. This pole, 2RS 1414, gives rise to the ``standard" 
$\Lambda(1405)$ resonance, clearly seen in the  $\pi\Sigma$ event distribution of figs.
\ref{fig:a4p}, \ref{fig:a4pnew} and \ref{fig:b4new}. Its width resulting from the pole
position\footnote{As it is well known, minus twice the imaginary part of the pole position is the width
of the resonance. Nonetheless, this is only so when the resonance is narrow and its difference to the
closest threshold is substantially larger than the width.} is
around 46 MeV. Their parameters, mass and width, are then in good agreement with those of 
the PDG \cite{pdg}.  The right most shape of this resonance, above the $\bar{K}N$ thresholds, does not
correspond to any pole in the 3RS plane and just corresponds to a cusp effect due to the opening of the 
$\bar{K}N$ thresholds. This behaviour is shown in the second panel of fig.\ref{fig:a4pnewpoles0}, where
 the square modulus of the I=0 $\bar{K}N$ S-wave is plotted. In this 
panel one can also observe a narrow pole between the $K^-p$ and $\bar{K}^0 n$ thresholds 
corresponding to a narrow I=1 pole to be discussed below. This pole appears in I=0 
because of isospin violation. In the 3RS
we find another pole at $1388-i\,17$ MeV that controls, modulo the cusp effect at the opening of the
$\bar{K}N$ thresholds, the size of the $\pi\Sigma$ I=0 amplitudes. This pole couples
much more weakly to $\bar{K}N$, and this is why it does not affect its shape in the physical sheet, 
see the second panel. These two latter poles, 2RS 1414 and 3RS 1388, are connected continuously
and, hence, reflect the same resonance, the $\Lambda(1405)$. As discussed above, before this resonance we also
have another one, peaked around the $\pi\Sigma$ threshold. In ref.\cite{teamL} the fact of having two nearby
poles around the nominal mass of the $ \Lambda(1405)$ was referred as the dynamics of the two
$\Lambda(1405)$. In our solution we still find two resonances,  but one of these
``$\Lambda(1405)$'' has moved to lower energies, and the two peaks can be distinguished now.
 We now consider the $\Lambda(1670)$ resonance \cite{pdg}, this is clearly visible in the
ref.\cite{nefkens} data on the reaction $K^- p\rightarrow \eta \Lambda$, as shown 
 in the eighth panel of figs. \ref{fig:a4p}, \ref{fig:a4pnew} and \ref{fig:b4new}. The left part of
 this resonance, before the opening of the $\eta \Lambda$ threshold, is driven by the pole in the 3RS
at $1676-i\,10$ MeV, while the right part, above the $\eta \Lambda$ threshold, is driven by the pole
in the 4RS at $1673-i\,13.5$ MeV. Both poles have similar values for mass and width although they
 are not the same, which is specially relevant in this case 
since the width is rather small, around 20 MeV, and because of  
the nearby position of the $\eta \Lambda$ threshold.  These poles have their largest couplings 
to the $\eta \Lambda$ and $K\Xi$ channels, around 2.1 and 5.3, respectively. We also warn that the
actual shape of the $\Lambda(1670)$ resonance can depend strongly on the process. For example, for the
square modulus of the $\bar{K}N$ or $\pi \Sigma$ elastic I=0 scattering amplitudes, the peak is 
shifted to higher energies,
towards 1.7 GeV, because of a strong distortion induced in these cases by the
$\eta \Lambda$ channel. For this channel the $\Lambda(1670)$ appears as a clean strong
 enhancement. However, its shape is an asymmetric distorted Breit-Wigner resonance
  because on the left of the $\eta
\Lambda$ threshold it has a width of around 20 MeV, from the 2RS 1676 pole, while on the right
 its width is larger, around 
26 MeV, from the 4RS 1673 pole. The poles 3RS 1676 and 4RS 1673 are connected continuously, 
as one would expect. They reflect, as discussed, the $\Lambda(1670)$ resonance.
In the 5RS there is another pole at $1825-i\,49$ MeV. This pole drives an increase in the
 I=0 amplitudes involving the $\pi\Sigma$, $\bar{K}N$ and $K\Xi$
 I=0 states to which it couples strongly, 2.3, 2.1 and 7.4, respectively, for a few tens of MeV
 before the $K\Xi$ threshold. The coupling to $\eta \Lambda$ is much
weaker, 0.9, and it does not give rise to any rapid movement for this case. This pole disappears
 in the 6RS, and for energies higher than the $K\Xi$ threshold one only has a remarkable
  cusp effect. This is accompanied by 
an important decrease in the values of the I=0 amplitudes for those to which the previous pole
strongly couples. See the third panel of fig.\ref{fig:a4pnewpoles0}, where the square modulus of the elastic I=0 
$K\Xi$ S-wave is shown. In the PDG \cite{pdg} there is an entry for the $\Lambda(1800)$ resonance with
values for its mass and width in correspondence with the pole position just given. However, we must
stress that its signal in any scattering amplitude is far from being that of a simple Breit-Wigner
because it appears just on top of the $K\Xi$ threshold (the distance to that is much smaller than 
 its width of around 100 MeV), and this pole appears in one sheet but not in the next one.
 E.g., its value for the width as minus twice the imaginary part of the pole position is not
 appropriate for this case to the right of the $K\Xi$ threshold. In the last panel of
fig.\ref{fig:a4pnewpoles0} we
show the square modulus of the same amplitude as in the third panel but 
 now, in addition, we also show the physical sheet, indicated by 0RS. It is remarkable how this sheet matches
along the real axis, because of continuity, first to the 5RS and then, after the $K\Xi$ threshold, to the 6RS.
To show this more clearly, we have plotted the amplitude in the isospin limit for this last panel,
 so that only one threshold is present. We give in table
 \ref{table:a4pnewpoles0} the pole positions and couplings of the discussed I=0 poles.

\begin{figure}[H]
\centerline{\epsfig{file=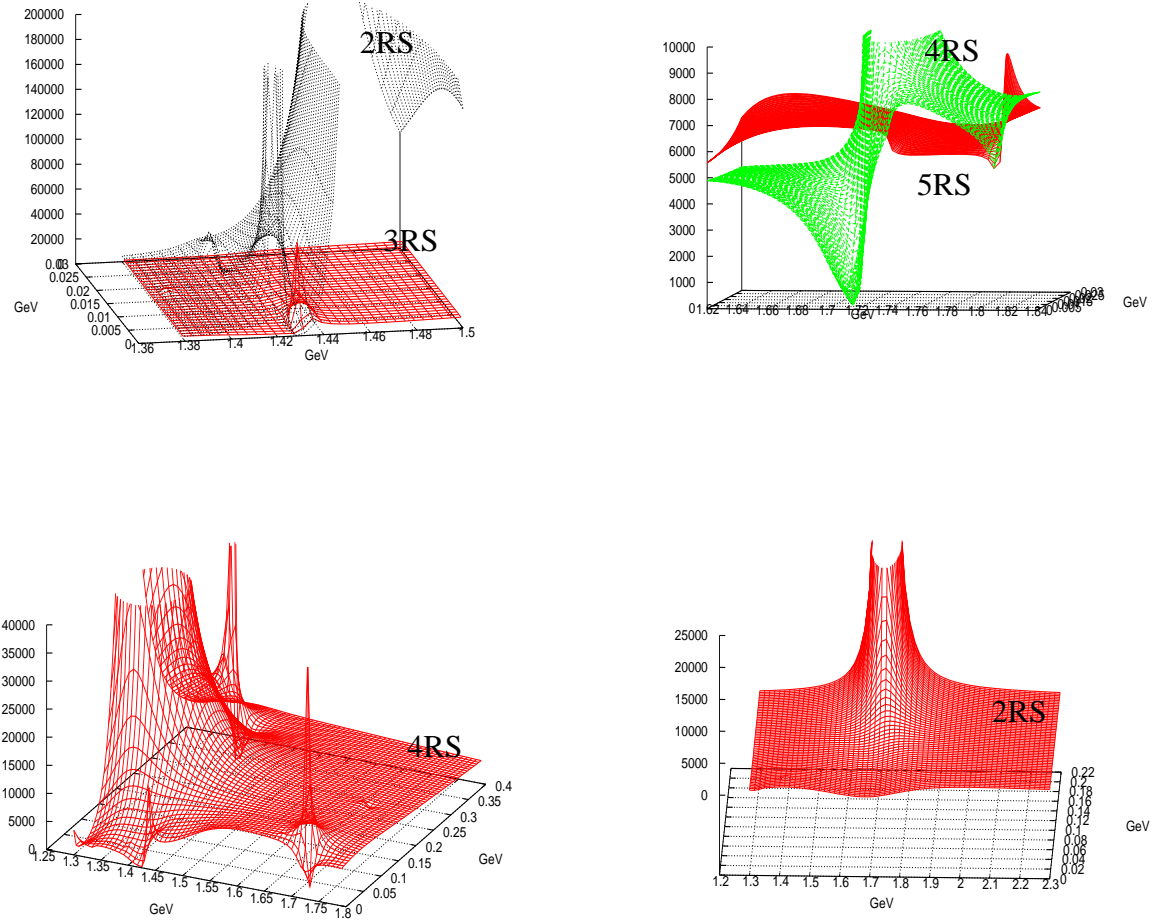,height=5.6in,width=6.5in,angle=0}}
\vspace{0.2cm}
\caption[pilf]{\protect \small
Some poles with I=1 for the fit I. From left to right and top to bottom, we have the square modulus 
 of the I=1  S-waves: $\pi\Sigma$, $\pi\Lambda$ and again of $\pi \Sigma$.
 The last panel corresponds to the square modulus of the 
I=2 $\pi\Sigma$ S-wave.
\label{fig:a4pnewpoles1}}
\end{figure}

\begin{table}[ht]
\begin{center}
\begin{tabular}{|r|r|r|r|r|r|r|r|r|r|}
\hline
Re(Pole) & -Im(Pole) &Sheet & & & & & & & \\
 $|g_{\pi\Lambda}|$ & $|g_{\pi\Sigma}|_0$ & $|g_{\pi\Sigma}|_1$ 
 & $|g_{\pi\Sigma}|_2$ & $|g_{\bar{K} N}|_0$ & $|g_{\bar{K} N}|_1$ & $|g_{\eta\Lambda}|$ 
 & $|g_{\eta\Sigma}|$ & $|g_{K\Xi}|_0$ & $|g_{K\Xi}|_1$\\
\hline
1425 & 6.5 & 2RS & & & & & & &\\
1.35 & 0.24 & 1.66 & 0.01 & 0.35 & 3.92 & 0.05 & 4.23 & 0.49 & 2.98 \\ 
\hline
1468 & 13 & 2RS & & & & & & &\\
2.80 & 0.16 & 5.96 & 0.02 & 0.23 & 8.74 & 0.04 & 10.66 & 0.19 & 2.48 \\ 
\hline
1433 & 3.7 & 3RS & & & & & & &\\
0.65 & 0.08 & 0.80 & 0.00 & 0.12 & 1.58 & 0.02 & 5.82 & 0.20 & 2.14 \\ 
\hline
1720 & 18 & 4RS & & & & & & &\\
1.82 & 0.02 & 1.21 & 0.00 & 0.02 & 0.95 & 0.02 & 6.78 & 0.05 & 5.31 \\ 
\hline
1769 & 96 & 6RS & & & & & & &\\
2.65 & 0.00 & 0.61 & 0.00 & 0.00 & 2.48 & 0.00 & 3.32 & 0.01 & 4.22 \\ 
\hline
1340 & 143 & 3-4RS  & & & & & & &\\
1.33 & 0.14 & 5.50 & 0.02 & 0.02 & 1.58 & 0.00 & 3.28 & 0.03 & 1.20\\
\hline
1395 & 311 & 3-4RS   & & & & & & &\\
2.08 & 0.01 & 1.49 & 0.01 & 0.00 & 1.24 & 0.00 & 7.63 & 0.01 & 3.97 \\
\hline
\end{tabular}
\caption{Fit I, I=1 Poles. The pole positions are given in MeV and the couplings in GeV. The 
couplings to the I=0, 2 channels are always close to zero.
\label{table:a4pnewpoles1}}
\end{center}
\end{table}

\item{{\bf I=1 Poles:}}
In the 2RS we find two narrow I=1 poles close to the $\bar{K}N$ thresholds.
 One located at $1425-i\,6.5$ MeV and the other at $1468-i\,13$ MeV. The first one has a $\pi\Sigma$
coupling of 1.7 while the other has a much stronger coupling of 6.0. They interfere destructively
around 1.42 GeV and there is a dip there, as shown in the first panel of fig.\ref{fig:a4pnewpoles1}, where 
the square modulus of the elastic I=1 $\pi \Sigma$ S-wave is plotted. Indeed, this is the only
observable signal in the square of the I=1 $\pi\Sigma$ elastic amplitude for the second pole, because
it disappears in the 3RS. Had the heavier pole not appeared  we would then have obtained a symmetric and
standard Breit-Wigner resonance shape for the pole at $1425-i\,6.5$ MeV. Instead, we find a sharp 
dip to the right of the pole position. This remarkable destructive
interference for the $\pi \Sigma$ and ${\bar K}N$ I=1 amplitudes below the ${\bar K}N$ threshold 
is due to the  large couplings of the pole in the 2RS
at $1468-i\,13$ MeV to $\pi \Sigma$ (6.0) and $\bar{K}N$ (8.7). This
effect is not so strong in the $\pi \Lambda$ case because its coupling  is 
smaller, 2.8. The 2RS pole $1425-i\,6.5$ MeV  evolves continuously in the 3RS to another pole 
at $1433-i\,3.7$ MeV. This pole drives the behaviour of the 
  I=1 amplitudes at the right of the $\bar{K}N$ threshold. These two poles, being connected, 
correspond to the same resonance. We see then that the  
behaviour of the I=1 amplitudes from around 1.4 up to 1.45 GeV is dominated by these three poles,
 giving rise to a pronounced peak structure in the amplitudes.
 Recently,  a signal for a resonance at
$1480\pm 15$ MeV and width of $60\pm 15$ MeV has been reported from the reaction $pp\to K^+ p Y^{0*}$ 
in ref.\cite{cosy}.
Maybe, one can invoke 
interference effects to account for the displacement of our peak around 1.43 GeV to somewhat higher energies,
  as observed by COSY. 
Other relevant pole for the I=1
 amplitudes is the one located at $1720-i\,18$ MeV on the 4RS. This pole is visible like a distorted
 bump in the $\pi \Lambda$ (1.82), $\pi \Sigma$ (1.21)  and $\bar{K}N$ (0.95) S-waves, however is a
 clean resonance signal for the not yet open $\eta \Sigma$ (6.78) and $K\Xi$ (5.31) channels. In the 5RS
this pole disappears and one only observes a  $\eta\Sigma$ cusp effect in some channels whose value
matches with the descending tail of the former pole in the 4RS. This enhancement corresponds to the
$\Sigma(1750)$ resonance of the PDG, in good agreement with its values of mass and width. We show this behaviour in
the second panel of fig.\ref{fig:a4pnewpoles1} by plotting the square modulus of the $\pi\Lambda$ S-wave.
 In the 6RS sheet we find a relevant
pole at $1769-i\,96$ MeV, with a width of around 200 MeV, which is  responsible for the size and 
the slow descending value of the I=1 amplitudes after the cusp around the $K\Xi$ thresholds. 
Hence, this pole
 cannot be observed directly as a bump in the physical axis. Regarding the $\Sigma(1620)$ \cite{pdg}, some
of the I=1 S-wave amplitudes show a broad bump after the $\bar{K}N$ threshold and before that
of the $\eta\Sigma$. These bumps, like the one shown in the second panel of
fig.\ref{fig:a4pnewpoles1} for the 5RS surface of the $\pi \Lambda$ elastic amplitude,
is due to the interference of multiple poles. Two
 of them have been already discussed, namely, the 3RS 1433 and 4RS 1720
 poles. In addition, there are other two broad poles  at $1340-i\,143$ and $1395-i\,311$ MeV,
  shown in the last two lines of 
 table \ref{table:a4pnewpoles1}, that appear 
 simultaneously in the 3RS and 4RS in the same positions\footnote{Notice that the $\eta \Lambda$
 channel is I=0 and this is why it does not appreciably modify the pole positions of these broad
 I=1 poles.}. These poles, because of their long
descending tails, control to a large extent the
  sizes of the I=1 amplitudes in this region . All this produces this interesting interference phenomenon of  soft bumps on the
physical axis as shown in the panel before the last one of fig.\ref{fig:a4pnewpoles1}. In this
panel the square modulus of the $\pi\Sigma$ elastic I=1 S-wave is drawn. For this case, the 
wider pole is not very relevant, however, it is so for the $\eta \Sigma$ channel because its
coupling to this channel is much larger than that of the lighter resonance at 1340 MeV. This is why
we have kept it. The pole $1340-i\,143$ MeV on the 3RS or 4RS sheets is connected continuously to the
 previous
2RS pole at $1468-i\,13$ MeV, hence, for this solution, the $\Sigma(1620)$ and $\Sigma(1480)$ 
 are clearly related. 
In table \ref{table:a4pnewpoles1} we collect the different I=1 poles for the fit I presented here.
 
\item{{\bf I=2 Pole:}} An exotic I=2 pole appears at $1722-i\,181$ MeV with a strong coupling to the
$\pi \Sigma$ I=2 channel (the only  one possible with I=2) of 6.18. This pole appears in the 2RS, 3RS,
4RS, 5RS and 6RS, since in all of them the momentum for the $\pi\Sigma$ has reversed sign and the only
channel that matters is the I=2 $\pi\Sigma$. This pole is actually the responsible for the I=2 $\pi
\Sigma$ size and gives rise to a soft and wide bump in the square  modulus of the I=2 amplitude, 
with a dip at 1.7 GeV due to a zero, see the last panel of fig.\ref{fig:a4pnewpoles0}, where the square of 
the modulus of the I=2 S-wave is shown. Because of this non uniform shape and for its rather large
 magnitude, of
the same order as that for the other isospin $\pi\Sigma$ channels, 
 one  can think of the possibility of detecting  such exotic state.
 \end{itemize}

\subsection{Fits II and ${\Op}$}

We consider here the pole content of the fits II and ${\Op}$, both fits are given in table
\ref{table:b4newvalues}. 
 Tables \ref{table:b4newpoles0} and 
\ref{table:b4newpoles1} correspond to the I=0, 1 poles of 
fit II, while the tables \ref{table:opnewpoles0} and
\ref{table:opnewpoles1} are the same for the ${\Op}$ fit. 
 
\begin{table}[ht]
\begin{center}
\begin{tabular}{|r|r|r|r|r|r|r|r|r|r|}
\hline
Re(Pole) & -Im(Pole) &Sheet & & & & & & & \\
 $|g_{\pi\Lambda}|$ & $|g_{\pi\Sigma}|_0$ & $|g_{\pi\Sigma}|_1$ 
 & $|g_{\pi\Sigma}|_2$ & $|g_{\bar{K} N}|_0$ & $|g_{\bar{K} N}|_1$ & $|g_{\eta\Lambda}|$ 
 & $|g_{\eta\Sigma}|$ & $|g_{K\Xi}|_0$ & $|g_{K\Xi}|_1$\\
\hline
1347 & 36 & 2RS & & & & & & &\\
0.02 & 6.48 & 0.12 & 0.02 & 2.60 & 0.10 & 1.42 & 0.01 & 0.32 & 0.07 \\
\hline
1427 & 18 & 2RS & & & & & & &\\
0.12 & 3.87 & 0.23 & 0.01 & 6.99 & 0.23 & 3.49 & 0.05 & 1.64 & 0.32 \\ 
\hline
1340 & 41 & 3RS & & & & & & &\\
0.07 & 5.92 & 0.08 & 0.01 & 0.62 & 0.08 & 2.33 & 0.01 & 0.75 & 0.04 \\ 
\hline
1667 & 8 & 4RS & & & & & & &\\
0.03 & 0.77 & 0.05 & 0.00 & 0.59 & 0.01 & 3.32& 0.02 & 12.17 & 0.08 \\ 
\hline
1667 & 8 & 5RS & & & & & & &\\
0.03 & 0.77 & 0.05 & 0.00 & 0.59 & 0.01 & 3.32& 0.03 & 12.17 & 0.06 \\ 
\hline
\end{tabular}
\caption{Fit II, I=0 Poles. The pole positions are given in MeV and the couplings in GeV. 
\label{table:b4newpoles0}}
\end{center}
\end{table}

For the I=0 poles of fit II we observe that the $\Lambda(1405)$ region is controlled by three poles, two in the
2RS and one in the 3RS. The first two control the shape before the $\bar{K}N$ threshold and the
latter after this threshold is open. The 2RS 1347 and 3RS 1340 poles couple very strongly to $\pi \Sigma$ 
while the 2RS 1427  pole couples very strongly to ${\bar K}N$. Hence, the resonance looks broader for the
$\pi\Sigma$ channel than for the $\bar{K}N$, since the resonance  for the former 
mostly corresponds to the broader poles 
while for the latter it is mainly due to the narrower one. As a further consequence, for $\bar{K}N$
 the resonance peak is
shifted to the right. This kind of behaviour is already described in detail in
ref.\cite{teamL}, where ${\Op}$ analyses were used. In addition, the two poles 2RS 1347 and 3RS 1340 
are connected when passing continuously from the 2RS to the 3RS. 
The $\Lambda(1670)$ resonance is described by the
two poles located in the same place both in the 4RS and 5RS, giving rise to a clean symmetric 
Breit-Wigner without asymmetries in the $\eta\Lambda$ threshold. Of course, these two poles are 
continuously connected. There is no signal for the $\Lambda(1800)$.

\begin{table}[ht]
\begin{center}
\begin{tabular}{|r|r|r|r|r|r|r|r|r|r|}
\hline
Re(Pole) & -Im(Pole) &Sheet & & & & & & & \\
 $|g_{\pi\Lambda}|$ & $|g_{\pi\Sigma}|_0$ & $|g_{\pi\Sigma}|_1$ 
 & $|g_{\pi\Sigma}|_2$ & $|g_{\bar{K} N}|_0$ & $|g_{\bar{K} N}|_1$ & $|g_{\eta\Lambda}|$ 
 & $|g_{\eta\Sigma}|$ & $|g_{K\Xi}|_0$ & $|g_{K\Xi}|_1$\\
\hline
1399 & 41 & 2RS & & & & & & &\\
1.49 & 0.09 & 5.58 & 0.01 & 0.13 & 4.92 & 0.08 & 0.73 & 0.03 & 4.99 \\ 
\hline
1424 & 3.6 & 2RS & & & & & & &\\
0.54 & 0.14 & 1.58 & 0.00 & 0.20 & 1.17 & 0.10 & 0.61 & 0.04 & 3.76 \\ 
\hline
1311 & 122 & 3-4RS & & & & & & &\\
2.63 & 0.05 & 4.61 & 0.01 & 0.02 & 3.44 & 0.02 & 0.60 & 0.03 & 3.60 \\ 
\hline
1426 & 3 & 3RS & & & & & & &\\
0.56 & 0.04 & 1.18 & 0.00 & 0.07 & 0.77 & 0.04 & 0.61 & 0.02 & 3.74 \\ 
\hline
\end{tabular}
\caption{Fit II, I=1 Poles. The pole positions are given in MeV and the couplings in GeV. 
\label{table:b4newpoles1}}
\end{center}
\end{table}

We now turn to the I=1 poles for fit II. 
The region below the $\bar{K}N$ threshold is governed by the 2RS poles at 
$1399-i\,41$ and $1424-i\,3.6$ MeV. The latter appears as a dip in the slope of the former since the
interference is destructive. Above the $\bar{K}N$ threshold one has again two poles, one broad and 
 the other narrow, located in the 3RS at $1311-i\,122$ and $1426-i\,3$ MeV, respectively.
  In this case, the latter appears as  a
clear peak in the slope of the former. The 2RS 1399 and 3RS 1311 poles are connected continuously
 and then
corresponds to the same resonance. The same happens for the 2RS 1424 and 3RS 1426 poles, 
as one would expect. All these poles give rise to a broad enhancement of the I=1 S-waves
 up to around 1.45 GeV. 
 We do not observe any signal for the $\Sigma(1750)$. Regarding the $\Sigma(1620)$, a few amplitudes, like that of
 $\bar{K}N$, exhibits an enhancement around 1.6 GeV. Nonetheless, there is no a clear pole
 structure driving this behaviour. The most remarkable facts are the presence of a broad pole on
 the 3-4RS at $1311-i\,122$ MeV, which controls, up to some
 extent, the size of the amplitudes in this region through 
 its falling tail, and the opening of the $K\Xi$ channel, that always produce a pronounced cusp effect, 
 sometimes like a bump others like a dip. Nonetheless, for fit I there were more amplitudes
manifesting bumps around 1.6 than now for fit II.

\begin{table}[ht]
\begin{center}
\begin{tabular}{|r|r|r|r|r|r|r|r|r|r|}
\hline
Re(Pole) & -Im(Pole) &Sheet & & & & & & & \\
 $|g_{\pi\Lambda}|$ & $|g_{\pi\Sigma}|_0$ & $|g_{\pi\Sigma}|_1$ 
 & $|g_{\pi\Sigma}|_2$ & $|g_{\bar{K} N}|_0$ & $|g_{\bar{K} N}|_1$ & $|g_{\eta\Lambda}|$ 
 & $|g_{\eta\Sigma}|$ & $|g_{K\Xi}|_0$ & $|g_{K\Xi}|_1$\\
\hline
1375 & 60 & 2RS & & & & & & &\\
0.04 & 7.55 & 0.07 & 0.02 & 5.45 & 0.07 & 2.20 & 0.02 & 0.78 & 0.07 \\
\hline
1429 & 22 & 2RS & & & & & & &\\
0.13 & 4.68 & 0.14 & 0.01 & 6.85 & 0.21 & 4.50 & 0.20 & 0.54 & 0.17 \\ 
\hline
1710 & 19 & 4RS & & & & & & &\\
0.05 & 0.27 & 0.03 & 0.00 & 1.73 & 0.03 & 2.56 & 0.05 & 10.75 & 0.09 \\ 
\hline
1710 & 19 & 5RS & & & & & & &\\
0.04& 0.27 & 0.04 & 0.00 & 1.73 & 0.05 & 2.56& 0.04 & 10.75 & 0.11 \\ 
\hline
\end{tabular}
\caption{Fit ${\Op}$, I=0 Poles. The pole positions are given in MeV and the couplings in GeV. 
\label{table:opnewpoles0}}
\end{center}
\end{table}

We now  consider the  ${\Op}$ fit. 
 The first I=0 pole occurs at $1375-i\,60$ MeV on the 2RS with a large coupling to $ \pi \Sigma$. This
pole interferes destructively with that at $1429-i\,22$ MeV and this is why, at this point, the elastic I=0
$\pi \Sigma$ S-wave has a dip. The latter pole couples very strongly with $\bar{K}N$ and it is seen as a
clear maximum in the $\bar{K}N\to \pi\Sigma$ partial wave. There is no pole at around 1.4 GeV in the 3RS
sheet and  the right tail of this resonance, after the  $\bar{K}N$ threshold, corresponds to a pronounced
cusp effect that falls down from the latter threshold. Thus, for all the fits I, II and ${\Op}$ we observe the
presence of a minimum in the I=0 amplitudes before 1.42 GeV, a maximum for the modulus of the I=0 
$\pi\Sigma$ S-wave before
such energy (for the fits I, II and ${\Op}$ the maximum  is located around 1.34, 1.36 and 1.38,
 in order), and a maximum  for the amplitudes involving the $\bar{K}N$ channel 
around its threshold. As discussed before, this is related to 
the so called dynamics of the two $\Lambda(1405)$ \cite{teamL}.
 The $\Lambda(1670)$ region is given by the poles 
 4RS and 5RS 1710, which have very similar properties and are continuously connected when passing from
one sheet to the other. These poles describe this resonance above and below the $\eta
\Sigma$ threshold, respectively. There is no signal for the $\Lambda(1800)$.

\begin{table}[ht]
\begin{center}
\begin{tabular}{|r|r|r|r|r|r|r|r|r|r|}
\hline
Re(Pole) & -Im(Pole) &Sheet & & & & & & & \\
 $|g_{\pi\Lambda}|$ & $|g_{\pi\Sigma}|_0$ & $|g_{\pi\Sigma}|_1$ 
 & $|g_{\pi\Sigma}|_2$ & $|g_{\bar{K} N}|_0$ & $|g_{\bar{K} N}|_1$ & $|g_{\eta\Lambda}|$ 
 & $|g_{\eta\Sigma}|$ & $|g_{K\Xi}|_0$ & $|g_{K\Xi}|_1$\\
\hline
1423 & 1.3 & 2RS & & & & & & &\\
0.52 & 0.12 & 0.72 & 0.00 & 0.18 & 1.28 & 0.11 & 1.49 & 0.02 & 2.30 \\ 
\hline
1494 & 116 & 2RS & & & & & & &\\
3.83 & 0.15 & 10.06 & 0.03 & 0.14 & 9.25 & 0.06 & 2.82 & 0.02 & 4.55 \\ 
\hline
1425 & 4.8 & 3RS & & & & & & &\\
0.93 & 0.05 & 1.22 & 0.00 & 0.11 & 1.63 & 0.07 & 2.71 & 0.01 & 3.43 \\ 
\hline
1796 & 69 & 4RS & & & & & & &\\
3.97 & 0.01 & 2.33 & 0.01 & 0.02 & 1.96 & 0.03 & 3.20 & 0.15 & 9.09 \\ 
\hline
1808 & 71 & 5RS & & & & & & &\\
3.71 & 0.01 & 1.94 & 0.00 & 0.01 & 2.36 & 0.02 & 1.15 & 0.11 & 8.16 \\ 
\hline
1350 & 254 & 3-4RS& & & & & & &\\
2.38 & 0.06 & 5.19 & 0.01 & 0.02 & 3.36 & 0.01 & 5.33 & 0.01 & 4.11 \\
\hline
\end{tabular}
\caption{Fit ${\Op}$, I=1 Poles. The pole positions are given in MeV and the couplings in GeV. 
\label{table:opnewpoles1}}
\end{center}
\end{table}

The narrow I=1 pole 2RS 1423 only has an appreciable coupling to the  $K\Xi$ channel, closed at such
energies,  and this is why it is so narrow. This pole appears as a dip in the increasing slope of the 
wide 2RS 1494 pole. In the 3RS one has another pole at $1424-i\,4.8$ MeV, continuously connected 
 to the 2RS 1423 one. As in fits I and II, all these
poles give rise to a clear resonance structure up to around 1.5 GeV, close to the nominal mass of 
the $\Sigma(1480)$ resonance, and overlapping when taking into account widths \cite{pdg}. In addition,
 the enhancement
corresponding to the $\Sigma(1750)$ \cite{pdg} appears again in this fit, as in fit I, and it corresponds
to the relatively wide 4RS $1796-i\,69$ and 5RS $1808-i\,71$ MeV poles, in the relevant sheets for 
energies below and above the $\eta\Sigma$
threshold, respectively. These two poles are connected continuously.
 The $\Sigma(1620)$ bumps disappear for this fit. Although
one has the mentioned resonances around the $\bar{K}N$ and $K\Xi$ thresholds, the amplitudes do
not display bumps between them. One also has in the 3-4RS a wide pole at $1350-i\,254$ MeV
that interferes appreciably in the physical axis with the previous resonances but, in this case, this
interference is negative, while in the fit I was positive. This pole is continuously connected
 to the 2RS 1494 pole.

We want to end this section with some SU(3) considerations. Without moving to the SU(3) limit, we
 calculate the interaction {\em kernels}, ${\cal T}$, appearing in eq.(\ref{u1}),
 for the SU(3) irreducible representation $\mathbf{1}$, $\mathbf{8s}$, 
$\mathbf{8a}$,  $\mathbf{10}$, $\mathbf{\overline{10}}$ and $\mathbf{27}$.
 These are the ones that originate
from the tensorial product $\mathbf{8}\otimes \mathbf{8}$ of the octets of baryons and
mesons. If one performs such an exercise, one realizes that for the fit I the $\mathbf{1}$, 
$\mathbf{8s}$, $\mathbf{8a}$ and $\mathbf{27}$ have attractive kernels. These representations can
accommodate four I=0 and three I=1 resonances, in  agreement with the resonance content
discussed above.  For the fit II,
one obtains attraction in the representations ${\mathbf 1}$, 
$\mathbf{8s}$, $\mathbf{8a}$ and $\mathbf{\overline{10}}$. In this case, the previous SU(3)
representations can accommodate three I=0 and 1 resonances, although only two I=1 
resonances finally appear.
 Similarly, for the ${\Op}$ fit
 one has attractive interaction in the representations ${\mathbf 1}$, 
$\mathbf{8s}$, $\mathbf{8a}$ and $\mathbf{\overline{10}}$. As before these representations can 
accommodate three I=0 and 1  resonances. This is in agreement with the fact that any of
 the latter two fits does not reproduce the $\Lambda(1800)$ resonance, while this is the case 
 for the fit I.
\section{Conclusions}

We have considered a wide set of experimental data that includes several $K^-p$ cross 
sections, namely, the elastic and charge exchange ones and productions of hyperons ($\pi^0 \Lambda$, 
$\pi^0\Sigma^0$, $\pi^-\Sigma^+$ and $\pi^+\Sigma^-$), the $\pi\Sigma$ event distribution 
from ref.\cite{hemingway}, the reaction $K^-p\to \pi^0\pi^0 \Sigma^0$ from ref.\cite{prakhov},
 including a $\pi^0\Sigma^0$ event distribution 
and the total cross section, the total cross section of
$K^-p\to\eta\Lambda$ \cite{nefkens}, three ratios $\gamma$, $R_c$ and $R_n$ of cross
sections at the threshold of $K^- p$ \cite{nowak,tovee}, the difference of the P- and
S-wave $\pi\Lambda$ phase shifts at the $\Xi^-$ mass \cite{hyperCP,e756}, $\delta_{\pi\Lambda}(\Xi)$, 
and the three quantities,
$m_p$, $a_{0+}^+$, $\sigma_{\pi N}$, calculated at ${\Opd}$ in baryon CHPT. Last, but not least,
we have paid special attention to the energy shift and width of the $\alpha$ line ($2p\to 1s$)
 of  kaonic hydrogen in connection to its recent and accurate measurement  by the DEAR Collaboration
 \cite{DEAR}. We have reviewed the fits of ref.\cite{opv}, including as well a new ${\Op}$ fit, and shown that
 they cannot reproduce the additional data considered in section \ref{sec:newfits}, that is, those from $K^-p\to \pi^0\pi^0\Sigma^0$
 \cite{prakhov} and from $K^- p\to\eta\Lambda$ \cite{nefkens}. These fits are given in table 
 \ref{table:a4pvalues}. We have then searched for new fits including from the beginning all the previous set 
 of data points. Several fits arise, namely, the ${\Opd}$ ones given in tables \ref{table:a4pnewvalues} and 
\ref{table:b4newvalues}, that reproduce most of the data. The only exceptions 
are $\delta_{\pi \Lambda}(\Xi)$ and the DEAR data, eq.(\ref{deardata}),
 that are not in agreement (within 
the present precision given by their last measurements from refs.\cite{hyperCP} and \cite{DEAR}, 
respectively)
with the ${\Opd}$ fits of table \ref{table:b4newvalues}. The ${\Op}$ fit given in that table does not reproduce in
addition the $K^-p\to\eta \Lambda$ total cross section. Remarkably, the ${\Opd}$ fits of table
\ref{table:a4pnewvalues} are able to reproduce the whole set of data and, taking into account the value for
$\sigma_{\pi N}$, around 50 MeV \cite{glsigma,pavan}, we consider as our main fit, the so called fit I, 
the last one in this table with 
$\sigma_{\pi N}=40^*$ MeV. Nonetheless, all of them give very similar results for the rest of quantities, as 
shown in fig.\ref{fig:a4pnew}. Indeed, the values of the fitted free parameters are very similar, as
shown in table \ref{table:a4pnewvalues}. The values of the parameters of our main fit are quite similar 
to those of the fit $A_4^+$ of ref.\cite{opv} as well. This was the first fit to provide a set of chiral
parameters leading to a simultaneous reproduction of the $K^- p$ scattering data considered in 
ref.\cite{opv} and the DEAR
 value on the width and shift of
kaonic hydrogen. In addition, we have analysed in detail the pole content of the fits 
I (main fit), II (the $40^*$ fit of table \ref{table:b4newvalues}), and ${\Op}$ from table 
\ref{table:b4newvalues}. 
 We have discussed with special care the pole content of fit I and shown how it reproduces the two 
 $\Lambda(1405)$ resonances, and the $\Lambda(1670)$, $\Lambda(1800)$, $\Sigma(1480)$, $\Sigma(1620)$
  and $\Sigma(1750)$ 
resonances, as called in the PDG. 
We have shown that there is no a one to one correspondence between poles and
resonances and that the pole structure of a resonance can indeed be very involved, particularly, as it is always
the case here, when there is a threshold in the nearby. One then must consider in detail
the  connection between
Riemann sheets, in order to disentangle which poles are responsible for such effects, and collect 
as the same resonance those poles that are connected when passing continuously from one sheet
 to the other. Regarding the
pole contents of fits II and ${\Op}$, the former does not contain any poles associated with
 the $\Lambda(1800)$
and the $\Sigma(1750)$ resonances, while the latter does not reproduce the $\Lambda(1800)$ nor the 
$\Sigma(1620)$ bumps. Finally,
 fit I gives rise to an exotic broad I=2 resonance that could be observed since its size is similar to
that of the other $\pi\Sigma$ isospin S-waves, its shape is non-uniform and is the only resonance present 
in I=2. Thus,
also from the point of view of spectroscopy, the fit I is the solution that fits better with the
present resonance content in S-wave strangeness $-1$ as given in the PDG \cite{pdg}, giving rise to all the
strangeness $-1$ S-wave resonances from the onset of the $\pi\Sigma$ channel 
up to energies above 1.8 GeV. We then conclude that
fit I is our preferred fit in view of its unique agreement with the scattering, atomic and spectroscopic
  experimental information.

\section*{Acknowledgements}
I would like to acknowledge fruitful common  work and discussions with 
my collaborators J. Prades and M. Verbeni. I also thank E. Oset for useful discussions 
and communications. Financial support by MEC (Spain)  grant 
 No. FPA2004-03470, the European Commission (EC)  
RTN Program Network ``EURIDICE'' 
 Contract No HPRN-CT-2002-00311 and the HadronPhysics I3
Project (EC)  Contract No RII3-CT-2004-506078 is acknowledged.



\begin{thebibliography}{99}
\bibitem{DEAR}{\sc G. Beer}  {\it et al.} {\sc [DEAR Collaboration]}, 
\newblock Measurement of the kaonic hydrogen X-ray spectrum,
\newblock {\em Phys. Rev. Lett.} {\bf 94}, 212302 (2005).\vs

\bibitem{sid}{\sc D. L. Sirghi and  F. Sirghi, (DEAR/SIDDHARTA Collaboration)}, 
 The physics of kaonic atoms at DAFNE, \\
 {\small http://www.lnf.infn.it/esperimenti/dear/DEAR\_RPR.pdf}\\
  {\sc C. Curceanu (DEAR/SIDDHARTA Collaboration)}, Precision
measurements of kaonic atoms at DAFNE,\\
 {\small www.tp2.ruhr-uni-bochum.de/vortraege/workshops/trento05/Petrascu.pdf}


\bibitem{akaki}{\sc U.-G. Mei{\ss}ner, U. Raha and A. Rusetsky}, 
Spectrum and decays of kaonic hydrogen,
{\em Eur. Phys. J.} {\bf C35}, 349 (2004).\vs

\bibitem{deser} {\sc S. Deser} {\it et al.},
\newblock Energy level displacements in $\pi$ mesonic atoms,
\newblock Phys. Rev. {\bf 96}, 774 (1954); 
\newblock {\sc T. L. Trueman},
\newblock Energy level shifts in atomic states of strongly-interacting
particles,
\newblock {\em Nucl. Phys.} {\bf 26}, 57 (1961).\vs




\bibitem{dalitz} {\sc R. H. Dalitz and S. F. Tuan}, Possible resonant state in pion-hyperon scattering, 
{\em Phys. Rev. Lett.} {\bf 2}, 425 (1959); 
The energy dependence of low energy $K^- p$ processes, 
 {\em Ann. Phys. {\bf 8},  100 (1959)}.

\bibitem{galileo} {\sc A.~D.~Martin, N.~M.~Queen and G.~Violini},
Consistency tests of the low-energy $K^- p$ parametrizations, {\em Nucl.\ Phys.}
 {\bf B10}, 481 (1969);  {\sc P.~M.~Gensini, R.~Hurtado and G.~Violini},
$\pi Y$ phenomenology from anti-$K$ $N$ scattering, {\em  PiN Newslett.}  {\bf 13}, 291 (1997).
  [arXiv:nucl-th/9709023];  {\sc B.~Di Claudio, A.~M.~Rodriguez-Vargas and G.~Violini},
The Adler-Weisberger sum rule and the sigma commutator for the kaon-proton system, 
 {\em Z.\ Phys.} {\bf C3}, 75 (1979).

\bibitem{martin}{\sc A.~D. Martin},  Kaon-nucleon parameters, {\em Nucl. Phys.} {\bf B179}, 33 (1979).\vs

\bibitem{juelich}
{\sc  R.~Buttgen, K.~Holinde and J.~Speth,}
 $K N$ scattering and meson exchange,  {\em Phys.  Lett.} {\bf B163}, 305 (1985); 
{\sc R.~Buettgen, K.~Holinde, A.~Mueller-Groeling, J.~Speth and P.~Wyborny},
A meson exchange model for the $K^+ N$ interaction,  {\em Nucl.\ Phys.}  {\bf A506}, 586 (1990); 
  {\sc A.~Mueller-Groeling, K.~Holinde and J.~Speth},
 $K^- N$ interaction in the meson exchange framework, {\em Nucl.\ Phys.} {\bf A513}, 557 (1990).


\bibitem{hamaie} {\sc T. Hamaie, M. Arima and K. Masutani}, Negative-parity hyperons 
 in the constituent quark model with meson-quark couplings, {\em Nucl. 
Phys.} {\bf A591}, 675 (1995)


\bibitem{landau}{\sc P.~J.~Fink, G.~He, R.~H.~Landau and J.~W.~Schnick},
 Bound states, resonances and poles in low-energy anti-$K N$ interaction
 models, {\em Phys.  Rev.}  {\bf C41}, 2720 (1990).



\bibitem{cloudy} {\sc E.~A.~Veit, B.~K.~Jennings, R.~C.~Barrett and A.~W.~Thomas},
 Kaon-nucleon scattering in an extended cloudy bag model,
  {\em Phys.\ Lett.} {\bf B 137}, 415 (1984).

\bibitem{schat}{\sc J.~L.~Goity, C.~L.~Schat and N.~N.~Scoccola,}
 Negative parity 70-plet baryon masses in the $1/N_c$ expansion,
  {\em Phys.  Rev.} {\bf D66}, 114014 (2002);   
  Masses of the $70^-$ baryons in large $N_c$ QCD,
 {\em Phys.\ Rev.\ Lett.}  {\bf 88},  102002 (2002).

\bibitem{kaisersiegel} {\sc N.~Kaiser, P.~B.~Siegel and W.~Weise},
  Chiral dynamics and the low-energy kaon-nucleon interaction, 
  {\em Nucl.\ Phys.}  {\bf A594}, 325 (1995).


\bibitem{npa} {\sc J.~A.~Oller and E.~Oset,}
  Chiral symmetry amplitudes in the S-wave isoscalar and isovector  channels
  and the $\sigma$, $f_0(980)$, $a_0(980)$ scalar mesons,
  {\em Nucl.\ Phys.}  {\bf A620}, 438 (1997); {\em (E)-ibid.}  {\bf A652}, 407 (1999).
  

\bibitem{oset}{\sc E. Oset and A. Ramos},
 Non perturbative chiral approach to s-wave anti-$K N$ interactions, 
 {\em Nucl. Phys.} {\bf A635}, 99 (1998).\vs


\bibitem{reportramos}{\sc J.~A.~Oller, E.~Oset and A.~Ramos,}
  Chiral unitary approach to meson-meson and meson-baryon interactions  and
  nuclear applications, {\em Prog.\ Part.\ Nucl.\ Phys.}  {\bf 45}, 157 (2000).

\bibitem{ollerm} {\sc J.~A. Oller and U.-G. Mei{\ss}ner}, 
Chiral dynamics in the presence of bound states: kaon-nucleon  interactions
revisited, {\em Phys. Lett.} {\bf B500}, 263 (2000).\vs

\bibitem{teamL} {\sc D. Jido, J.~A. Oller, E. Oset, A. Ramos and U.-G. Mei{\ss}ner}, 
 Chiral dynamics of the two $\Lambda(1405)$ states, {\em Nucl. Phys.} {\bf A725}, 181 (2003).\vs


\bibitem{lutznieves} {\sc C. Garcia-Recio, M.~F.~M. Lutz and J. Nieves},
 Quark mass dependence of s-wave baryon resonances,
  {\em Phys.\ Lett.}  {\bf B582}, 49 (2004).


\bibitem{bura}
  {\sc B.~Borasoy, R.~Nissler and W.~Weise},
Kaonic hydrogen and $K^- p$ scattering,
{\em Phys. Rev. Lett.} {\bf 94}, 213401 (2005);  
Chiral dynamics of kaon nucleon interactions, revisited,
  {\em Eur.  Phys.  J.} {\bf A25}, 79 (2005); 
  Comment on `Surprises in threshold antikaon-nucleon physics',{\em arXiv:hep-ph/0512279}.

\bibitem{opv}{\sc J.~A. Oller, J. Prades} and {\sc M. Verbeni}, 
\newblock Surprises in threshold antikaon
nucleon physics,
\newblock {\em Phys. Rev. Lett.} {\bf 95}, 172502 (2005).\vs 

\bibitem{reply}{\sc J.~A.~Oller, J.~Prades and M.~Verbeni,}
  Reply to comment on `Surprises in threshold antikaon-nucleon physics',
  {\em arXiv:hep-ph/0601109.}

\bibitem{nefkens}{\sc A.~Starostin} {\it et al.} {\sc [Crystal Ball Collaboration]},
\newblock  Measurement of $K^- p \to \eta \Lambda$ near threshold,
\newblock  {\em Phys.\ Rev.\ C} {\bf 64}, 055205 (2001).\vs


\bibitem{prakhov} {\sc S.~Prakhov} {\it et al.}  {\sc [Crystal Ball Collaboration]},
\newblock $K^- p \to \pi^0 \pi^0 \Sigma^0$ at $p_{K^-}=514$-MeV/c to 750-MeV/c
 and comparison with other $\pi^0 \pi^0$ production,
\newblock {\em Phys.\ Rev.\ } {\bf C70}, 034605 (2004).\vs


\bibitem{magas}{\sc V.~K. Magas, E. Oset and A. Ramos,}
 Evidence for the two pole structure of the $\Lambda(1405)$ resonance,
 {\em Phys.\ Rev.\ Lett. } {\bf 95}, 052301 (2005).\vs

\bibitem{kaplan} {\sc D.~B.~Kaplan and A.~E.~Nelson},
 Strange goings on in dense nucleonic matter,  {\em Phys. Lett.}  {\bf B175}, 57 (1986).

\bibitem{brown} {\sc G.~Q.~Li, C.~H.~Lee and G.~E.~Brown},
 Kaons in dense matter, kaon production in heavy-ion collisions, and  kaon
condensation in neutron stars,  {\em Nucl.\ Phys.}  {\bf A625}, 372 (1997).

\bibitem{pons} {\sc J.~A.~Pons, S.~Reddy, P.~J.~Ellis, M.~Prakash and J.~M.~Lattimer},
 Kaon condensation in proton-neutron star matter, {\em Phys.\ Rev.}  {\bf C62}, 035803 (2000).

\bibitem{starts}  {\sc A.~Sedrakian},
 The physics of dense hadronic matter and compact stars, {\em arXiv:nucl-th/0601086}.



\bibitem{kaos} 
  {\sc W.~Scheinast {\it et al.}  [KaoS Collaboration]},
First observation of in-medium effects on phase space distributions of
 antikaons measured in proton-nucleus collisions, {\em  arXiv:nucl-ex/0512028}; 
   {\sc F.~Laue {\it et al.}  [KaoS Collaboration]},
 Medium effects in kaon and antikaon production in nuclear collisions at
subthreshold beam energies,  {\em Phys.\ Rev.\ Lett.}  {\bf 82}, 1640 (1999).
  
\bibitem{fuchs} {\sc C.~Fuchs},
 Kaon production in heavy ion reactions at intermediate energies,
  {\em Prog.\ Part.\ Nucl.\ Phys.}  {\bf 56}, 1 (2006).


\bibitem{ramoset} {\sc C.~Garcia-Recio, E.~Oset, A.~Ramos and J.~Nieves},
Non-localities and Fermi motion corrections in $K^-$ atoms, {\em Nucl.\ Phys.}  {\bf A703}, 271
 (2002);   {\sc A.~Baca, C.~Garcia-Recio and J.~Nieves},
Deeply bound levels in kaonic atoms, {\em Nucl.\ Phys.} {\bf A673}, 335 (2000).


\bibitem{gzero} {\sc D.~S. Armstrong {\it et al.}  (G0 Collaboration)}, 
Strange quark contributions to parity-violating asymmetries in the  forward
  G0 electron-proton scattering experiment 
{\em Phys. Rev. Lett.} {\bf 95}, 092001 (2005). 
 {\sc K.~A. Aniol {\it et al.}  (HAPPEX Collaboration)}, 
 Parity-violating electron scattering from $^4He$ and the strange electric form factor
of the nucleon, {\em Phys. Rev. Lett.} {\bf 96}, 022003 (2006); 
{\sc D.~T. Spayde {\it et al.} (SAMPLE Collaboration)}, 
The strange quark contribution to the proton's magnetic moment,
 {\em Phys. Lett.} {\bf B583}, 79 (2004); 
{\sc F.~E. Maas {\it et al.} (PVA4 Collaboration)}, 
Evidence for strange quark contributions to the nucleon's form factors  at
 $Q^2 = 0.108-(GeV/c)^2$, 
{\em Phys. Rev. Lett.} {\bf 94}, 152001  (2005).

\bibitem{pavan} {\sc M.~M. Pavan {\it et al.}}, 
The pion nucleon $\Sigma$ term is definitely large: Results from a GWU
 analysis of pi N scattering data, {\em PiN Newslett.} 
{\bf 16}, 110 (2002). 


\bibitem{fisica}{\sc S.~Weinberg,} 
Phenomenological Lagrangians,  {\em Physica} {\bf A96}, 327 (1979).
  
\bibitem{gl}{\sc J.~Gasser and H.~Leutwyler,}
  Chiral perturbation theory to one loop,  {\em Ann. Phys.}  {\bf 158}, 142 (1984); 
    Chiral perturbation theory: expansions in the mass of the strange quark, 
    {\em Nucl.\ Phys.} {\bf B250}, 465 (1985). 
  
\bibitem{ureport} {\sc U.-G. Mei{\ss}ner},  
    Recent developments in chiral perturbation theory, 
    {\em   Rept.\ Prog.\ Phys.}  {\bf 56}, 903 (1993).

\bibitem{u2} {\sc U.-G. Mei{\ss}ner},   Chiral QCD: baryon dynamics, 
{\em Shifman, M. (ed.): At the frontier of particle physics, vol. 1* 417-505, 
World Scientific,  arXiv:hep-ph/0007092}.
  
\bibitem{kaiserport} {\sc V.~Bernard, N.~Kaiser and U.-G.~Mei{\ss}ner,}
 Chiral dynamics in nucleons and nuclei,  {\em Int.\ J.\ Mod.\ Phys.}
   {\bf E4}, 193 (1995).
  
\bibitem{preport}{\sc A.~Pich},
  Chiral perturbation theory,
  {\em Rept.\ Prog.\ Phys.}  {\bf 58}, 563 (1995).
  
\bibitem{eckerport} {\sc G.~Ecker,} Chiral perturbation theory,
  {\em Prog.\ Part.\ Nucl.\ Phys.}  {\bf 35}, 1 (1995).
  
\bibitem{gasserb}{\sc J.~Gasser, M.~E.~Sainio and A.~Svarc,}
 Nucleons with chiral loops,
  {\em Nucl.\ Phys.}  {\bf B307}, 779 (1988).

\bibitem{weinn} {\sc S.~Weinberg,}
 Effective chiral Lagrangians for nucleon-pion interactions and nuclear
  forces,  {\em Nucl.\ Phys.}  {\bf B363}, 3 (1991).


\bibitem{kaiserkn} {\sc N.~Kaiser,} Chiral corrections to kaon-nucleon
 scattering lengths,  {\em Phys.\ Rev.} {\bf C64}, 045204 (2001).

\bibitem{boram} {\sc B. Borasoy and U.-G. Mei{\ss}ner}, 
Chiral expansion of baryon masses and sigma-terms,
 {\em Ann. Phys.} {\bf 254}, 192 (1997).\vs

\bibitem{steinm} {\sc U.-G.~Mei{\ss}ner and S.~Steininger},
 Baryon magnetic moments in chiral perturbation theory,
  {\em Nucl.\ Phys.}  {\bf B499}, 349 (1997).

\bibitem{iam}  {\sc J.~A.~Oller, E.~Oset and J.~R.~Pelaez,}
  Non-perturbative approach to effective chiral Lagrangians and meson
  interactions,
  {\em Phys.\ Rev.\ Lett.}  {\bf 80}, 3452 (1998); 
  Meson-meson and meson-baryon interactions in a chiral non-perturbative
  approach,  {\em Phys.\ Rev.}  {\bf D59}, 074001 (1999); {\em  (E)-ibid.}  
  {\bf D60} (1999) 099906.


\bibitem{nd}
  {\sc J.~A.~Oller and E.~Oset,}
  N/D description of two meson amplitudes and chiral symmetry,
  {\em Phys.\ Rev.}  {\bf D60}, 074023 (1999).


\bibitem{pin}{\sc U.-G.~Mei{\ss}ner and J.~A.~Oller,}
  Chiral unitary meson-baryon dynamics in the presence of resonances:
  Elastic pion-nucleon scattering,
  {\em Nucl.\ Phys.}  {\bf A673}, 311 (2000).

\bibitem{hyper} {\sc P.~G. Ratcliffe}, SU(3) breaking in hyperon beta decays: 
A prediction for  $\Xi^0 \to \Sigma^+ e^-  \bar{\nu}$, 
{\em Phys. Rev.} {\bf D59}, 014038 (1999).
%

\bibitem{opv2} {\sc J.~A. Oller, J. Prades and M. Verbeni}, in preparation.\vs



\bibitem{26plb}{\sc W.~E. Humphrey and R.~R. Ross}, Low-energy interactions of $K^-$-mesons
in hydrogen, 
{\em Phys. Rev.} {\bf 127}, 1305 (1962). 

\bibitem{27plb}{\sc J.K. Kim}, 
Low-energy $K^-p$ interaction and interpretation of the $1405-$MeV $Y^{0*}$ resonance as 
a $\bar{K}N$ bound state, {\em Phys. Rev. Lett.} {\bf 14}, 29 (1965).

\bibitem{28plb} {\sc M. Sakitt {\it et al.}}, 
Low-energy $K^-$-meson interactions in hydrogen, 
{\em Phys. Rev.} {\bf 139}, B719 (1965).

\bibitem{31plb} {\sc J. Ciborowski {\it et al.}}, Kaon scattering and charged sigma hyperon
production 
in $K^- p$ interactions below 300 MeV/c, {\em J. Phys.} {\bf G8}, 13 (1982).

\bibitem{29plb} {\sc W. Kittel, G. Otter and I. Wacek}, The $K^- p$  charge exchange interactions
 at low energies and scattering lengths determination, {\em Phys. Lett.} {\bf 21}, 349 (1966). 

\bibitem{30plb}{\sc D.~Evans {\it et al.}}, 
Charge exchange scattering in $K^- p$ interactions below 300-MeV/c,
  {\em J.\ Phys.}  {\bf G9}, 885 (1983).

\bibitem{nowak}{\sc R.J. Nowak {\it et al.}}, 
Charged $\Sigma$ hyperon production by $K^-$-meson interactions at rest, 
 {\em Nucl. Phys.} {\bf B139}, 61 (1978).\vs
 
\bibitem{tovee}{\sc D. Tovee {\it et al.}}, 
Some properties of the charged $\Sigma$ hyperons, {\em Nucl. Phys.} {\bf B33}, 493 (1971).\vs


\bibitem{hemingway} {\sc R.J. Hemmingway},
 Production of $\Lambda (1405)$ in $K^- p$ reactions at 4.2-GeV/c, 
 {\em Nucl. Phys.} {\bf B253}, 742 (1985).\vs


\bibitem{pdg}{\sc S.~Eidelman {\it et al.}  [Particle Data Group]},
 Review of particle physics,
  {\em Phys.\ Lett.}  {\bf B592}, 1  (2004).

\bibitem{cosy}{\sc I.~Zychor {\it et al.}},
 Evidence for an excited hyperon state in $p p \to$ $p K^+ Y^{*0}$,
  {\em Phys.\ Rev.\ Lett.}  {\bf 96}, 012002 (2006).
  
\bibitem{ben} {\sc E.~Oset, A.~Ramos and C.~Bennhold},
 Low lying $S=-1$ excited baryons and chiral symmetry, 
 {\em  Phys.  Lett.} {\bf B527}, 99 (2002); (E)-ibid.  {\bf B530}, 260 (2002).

\bibitem{lutzkolo} {\sc M.~F.~M. Lutz and E.~E. Kolomeitsev,}
 Baryon resonances from chiral coupled-channel dynamics,
  {\em Nucl.  Phys.}  {\bf A755}, 29 (2005).
  
\bibitem{ito}{\sc M. Iwasaki} {\it et al.},
\newblock Observation of the kaonic hydrogen $K(\alpha)$ x-ray,
\newblock Phys. Rev. Lett. {\bf 78}, 3067 (1997);
\newblock {\sc T.M. Ito} {\it et al.}, Observation of kaonic hydrogen atom x-rays, 
\newblock {\em Phys. Rev.} {\bf C58}, 2366 (1998).\vs

\bibitem{ulf}{\sc V. Bernard, N. Kaiser and U.-G. Mei{\ss}ner}, 
Chiral corrections to the S-wave pion-nucleon scattering lengths, 
{\em Phys. Lett.} {\bf B309}, 421 (1993).


\bibitem{gasser2} {\sc J. Gasser}, 
\newblock Hadron masses and sigma commutator in the light of chiral perturbation theory,
\newblock {\em Ann. Phys.} {\bf 136}, 62 (1981).

\bibitem{glsigma}{\sc J. Gasser, H. Leutwyler and M. Sainio}, Sigma term update,
  {\em Phys.\ Lett.}  {\bf B253}, 252 (1991).


\bibitem{frink} {\sc M. Frink, U.-G. Mei{\ss}ner and I. Scheller},
\newblock Baryon masses, chiral extrapolations, and all that,
\newblock {\em Eur. Phys. J.} {\bf A24}, 395 (2005);  {\sc M. Frink and U.-G. Mei{\ss}ner},  
Chiral extrapolations of baryon masses for unquenched three-flavor  lattice
simulations,  JHEP {\bf 0407} (2004) 028.\vs
  
\bibitem{schroder}{\sc H.~C. Schr\"oder {\it et al.}},
 Determination of the $\pi N$ scattering lengths from pionic hydrogen, 
 {\em Phys. Lett.} {\bf B469}, 25 (1999).\vs

\bibitem{baxter}{\sc D.~F.~Baxter {\it et al.}},
 A study of neutral final states in $K^-p$ interactions in the range from 690
 to 934 MeV/C,  {\em Nucl.\ Phys.}  {\bf B67}, 125 (1973).
  
\bibitem{minuit}F. James, {\em Minuit reference manual} {\bf D506} (1994).  
  
  
\bibitem{hyperCP} {\sc M. Huang} {\it et al.} {\sc [HyperCP Collaboration]}, 
\newblock New measurement of $\Xi^- \to \Lambda \pi^-$ decay parameters,
\newblock {\em Phys. Rev. Lett.} {\bf 93}, 011802 (2004).
 
\bibitem{e756} {\sc A. Chakravorty} {\it et al.} {\sc [E756 Collaboration]}, 
\newblock Measurement of decay parameters for $\Xi^- \to \Lambda \pi^-$ decay,
\newblock {\em Phys. Rev. Lett.} {\bf 91}, 031601 (2003).\vs

\bibitem{valencia} {\sc J. Tandean, A.W. Thomas and G. Valencia},
\newblock Can the $\Lambda \pi$ scattering phase shifts be large?, 
\newblock {\em Phys. Rev.} {\bf D64}, 014005 (2001).

\bibitem{lebed} {\sc R.~F. Lebed and M.~A. Luty,} Baryon masses at second order in chiral
 perturbation theory,  {\em Phys.\ Lett.\ } {\bf B329}, 479 (1994).\vs
 


\bibitem{holsteinm0}{\sc J.~F. Donoghue, B.~R. Holstein and B. Borasoy},
SU(3) baryon chiral perturbation theory and long distance
regularization,  {\em Phys.\ Rev.\ } {\bf D59}, 036002 (1999).\vs
  


\bibitem{sherer} {\sc B.~C. Lehnhart, J. Gegelia and S. Scherer}, 
Baryon masses and nucleon sigma terms in manifestly Lorentz-invariant
 baryon chiral perturbation theory, {\em  J.\ Phys.\ G} {\bf 31}, 89 (2005).\vs
  
 
\bibitem{lutzm0}  {\sc A. Semke and M.~F.~M. Lutz}, 
Baryon self energies in the chiral loop expansion, 
  {\em arXiv:nucl-th/0511061.}\vs


\bibitem{manley} {\sc D.~M.~Manley {\it et al.}}, 
Properties of the $\Lambda(1670)\,(1/2)^-$ resonance,
{\em Phys.\ Rev.\ Lett.}  {\bf 88}, 012002 (2002).



\bibitem{nieves}{\sc C. Garcia-Recio, J. Nieves, E. Ruiz Arriola and M.~J. Vicente Vacas},
$S = -1$ meson-baryon unitarized coupled channel chiral perturbation theory
and the $S_{(01)}$ $-\Lambda(1405)$ and $- \Lambda(1670)$ resonances,
  {\em Phys.\ Rev. } {\bf D67}, 076009 (2003).\vs


\bibitem{vicente} {\sc T. Hyodo, A. Hosaka, E. Oset, A. Ramos and M.J. Vicente Vacas}, 
$\Lambda(1405)$ production in the $\pi^- p \to K^0 \pi \Sigma$ reaction,
{\em Phys. Rev.} {\bf C70}, 034605 (2004).\vs


\bibitem{ollerm2} {\sc U.-G. Mei{\ss}ner and J.~A. Oller,}
\newblock  The S-wave $\Lambda \pi$ phase shift is not large,
\newblock {\em Phys. Rev.} {\bf D64}, 014006 (2001).\vs



\end{thebibliography}
\end{document}